\documentclass[fleqn,usenatbib]{mnras}

\usepackage{newtxtext,newtxmath}
\usepackage[T1]{fontenc}

\DeclareRobustCommand{\VAN}[3]{#2}
\let\VANthebibliography\thebibliography
\def\thebibliography{\DeclareRobustCommand{\VAN}[3]{##3}\VANthebibliography}
\usepackage{graphicx}	% Including figure files
\usepackage{amsmath}	% Advanced maths commands

\usepackage{algorithm}
\usepackage{algorithmicx}
\usepackage{dsfont}
\usepackage{bbm}
\usepackage[noend]{algpseudocode}
\usepackage{placeins}
\usepackage{subcaption}
\usepackage{lineno}
\usepackage{orcidlink} %enable \orcidlink{} (package added by Genaro)
\usepackage{appendix}

\newcommand{\ticone}{TIC\,13955147}
\newcommand{\tictwo}{TIC\,269797536}
\newcommand{\ticfour}{TIC\,441420236}
\newcommand{\aumic}{AU\,Mic}

\usepackage{graphicx}	% Including figure files
\usepackage{amsmath}	% Advanced maths commands
\title[Detecting stellar flares using conditional volatility]{Detecting stellar flares in the presence of a deterministic trend and stochastic volatility}

\author[Q. Wang et al.]{
Qiyuan Wang$^{1}$,
Giovanni Motta$^{2}$,
Genaro Sucarrat$^{3}$, and 
Vinay L.\ Kashyap$^{4}$\thanks{E-mail: vkashyap@cfa.harvard.edu}
\\
$^{1}$Department of Statistics, Texas A\&M University | 
155 Ireland St, College Station, TX 77843, USA\\
$^{2}$Data Science Institute, Columbia University | 550 W 120th St, New York, NY 10027, USA\\
$^{3}$Department of Economics, BI Norwegian Business School | Nydalsveien 37, 0484 Oslo, Norway {\orcidlink{0000-0002-8433-837X}}\\
$^{4}$Center for Astrophysics, Harvard \& Smithsonian $|$ 60 Garden St, Cambridge MA 02138, USA\\
}

\date{Accepted 2025 December 17. Received 2025 November 29; in original form 2025 June 21}

\pubyear{2026}

\begin{document}
\label{firstpage}
\pagerange{\pageref{firstpage}--\pageref{lastpage}}
\maketitle

\begin{abstract}
{We develop a new and powerful method to analyze time series to rigorously detect flares in the presence of an irregularly oscillatory baseline, and apply it to stellar light curves observed with TESS. First, we remove the underlying non-stochastic trend using a time-varying amplitude harmonic model.  We then model the stochastic component of the light curves in a manner analogous to financial time series, as an ARMA+GARCH process, allowing us to detect and characterize impulsive flares as large deviations inconsistent with the correlation structure in the light curve.  We apply the method to exemplar light curves from \ticone\ (a G5V eruptive variable), \tictwo\ (an M4 high-proper motion star), and \ticfour\ (\aumic, an active dMe flare star), detecting up to 145, 460, and 403 flares respectively, at rates ranging from ${\approx}0.4-8.5$~day$^{-1}$ over different sectors and under different detection thresholds.  We detect flares down to amplitudes of $0.03$\%, $0.29$\%, and $0.007$\% of the bolometric luminosity for each star respectively.  We model the distributions of flare energies and peak fluxes as power-laws, and find that the solar-like star exhibits values similar to that on the Sun ($\alpha_{E,P}\approx1.85,2.36$), while for the less- and highly-active low-mass stars $\alpha_{E,P}>2$ and $<2$ respectively.}
\end{abstract}


%%%%%%%%%%%%%%%%%%%%%%%%%%%%%%%%%%%%%%%%%%%%%%%%%%

%%%%%%%%%%%%%%%%% BODY OF PAPER %%%%%%%%%%%%%%%%%%

\section{Introduction}\label{sec:intro}

Flares occur when magnetic energy stored in twisted flux tubes is suddenly released into the hot plasma of the stellar corona.  However, while flare evolution is well studied \cite[see, e.g.,][]{2004A&A...416..713G,2014LRSP...11....4R,Testa2022}, flare onset, while attributable in principle to magnetic reconnection, is poorly understood \cite[][]{Zweibel_Yamada_2009}.  Flares are thought to be realized from a self-organized critical (SOC) process \citep[][]{1991ApJ...380L..89L} and a useful diagnostic to its nature likely lies in the distribution of the energies of individual flares, which has been established to follow a power-law form over several orders of magnitude \citep[see, e.g.,][]{2016SSRv..198...47A}, on both the Sun and other stars \citep[][]{2000ApJ...541..396A,2002ApJ...580.1118K,2004A&ARv..12...71G,Feinstein2022}\footnote{The number of flares $N$ at or greater than a given energy $E$ is usually described as 
$$N(>E) \propto E^{-\alpha+1} \,$$
with $\alpha\approx{1.8}$ on the Sun, but ranges to $\alpha>2$ on some active stars.}.  Detecting flares and determining their distributions in particular ensembles of stars is helpful to characterize their activity levels and the habitability of planets around them \citep{2025arXiv250515451G}.

An important differentiator between the Sun and other stars is in the flare energies accessible to observation: stellar flares are typically more energetic than seen on the Sun, which is a consequence of stellar flares occurring at greater distances and thus weaker flares are harder to detect.  It is thus crucial to improve the detectability of stellar flares so that like-to-like comparisons may be made with solar flares.  Several groups have worked to develop a varied set of methods to detect flares, in TESS light curves in particular, ranging from using outliers from running medians \citep{2020AJ....159...60G}, to Gaussian Process regression and decay modeling \citep{2022A&A...661A.148C}, to Convolutional Neural Networks \citep{2020AJ....160..219F}.

Here we develop a novel method based on stochastic modeling of the light curves to detect flare occurrences in stars in order to gain insight into their distributions.
Our goal here is to define the detection process, with the future aim of analyzing the large dataset of light curves obtained with TESS to characterize the different stars \citep[as in, e.g.,][]{2025arXiv250515451G}.
We study several white light flares from the Transiting Exoplanet Survey Satellite (TESS), and model them via a GARCH process. The GARCH process has been applied extensively in econometrics to understand stock price fluctuations, and our preliminary findings show that stellar flares can also be described by the same mathematics. We study the so-called conditional volatility $\sigma_t$, and find that it is a good guide to flare occurrence. This is a remarkable new result, and promises to open new avenues of exploration, to understand the precise nature of flare onset.  

\begin{figure}
\includegraphics[width=\columnwidth]{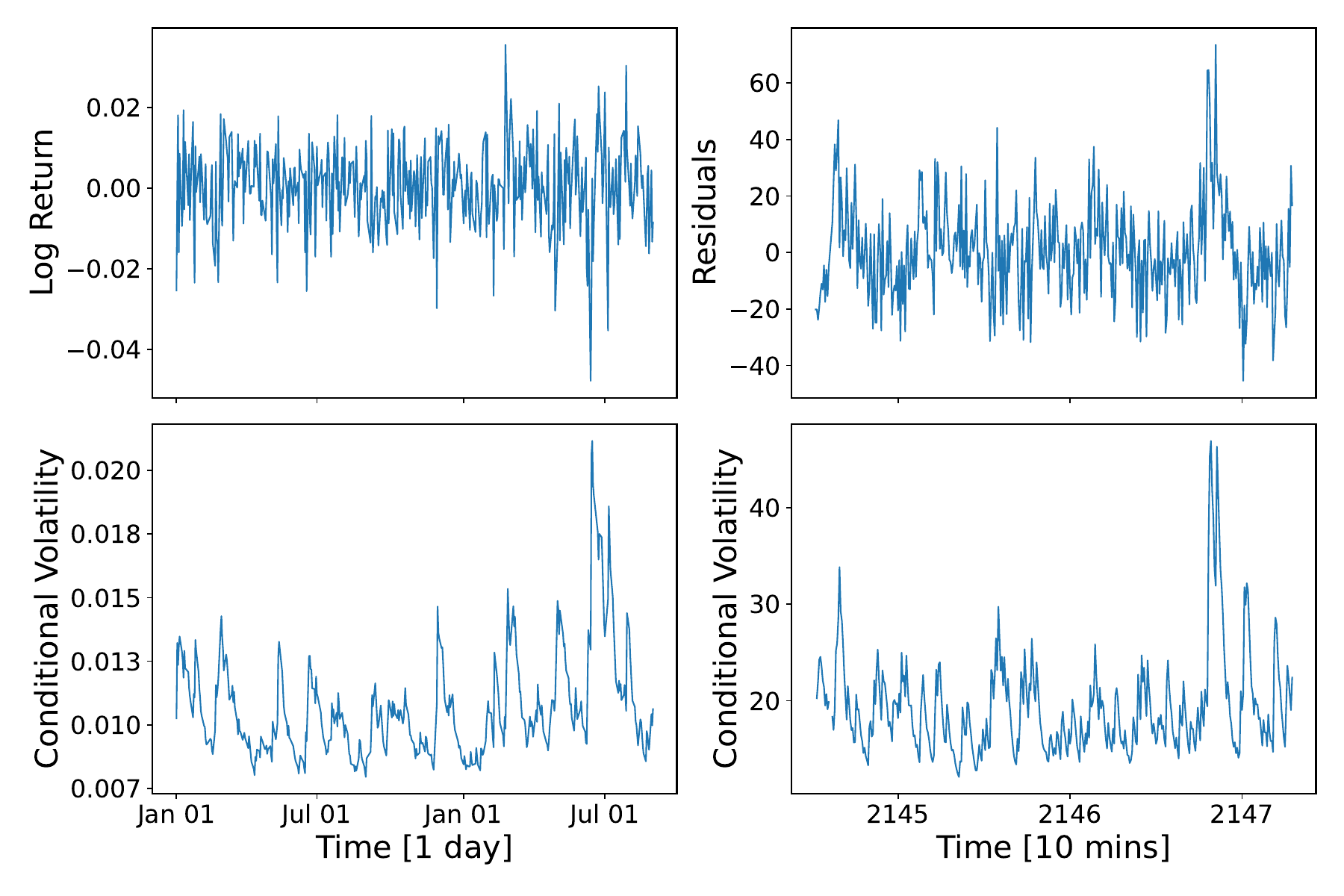}
    \caption{
    Illustration of the similarity between financial returns and astronomical time series.  The time series of the log returns of the Dow Jones Utility Average (sampled daily) during the years 2021 and 2022 (DJUA; {\sl top left}) bears remarkable similarity to the TESS light curve (at a 10~minute cadence) of the star \ticone-s0031 after removing the harmonic baseline ({\sl top right}).  Both time series have $\approx$500 points.  The fitted {\sl conditional volatility} for both time series are shown in the corresponding lower panels, with the DJUA ({\sl bottom left}) showing variations that are similar to those seen for the star ({\sl bottom right}).  This similarity motivates our adoption of techniques developed for financial data analysis to TESS light curves.
    }
    \label{fig:compar}
\end{figure}

For more than forty years now, following the Nobel Prize winning article by \citet{Engle1982} on the Autoregressive Conditional Heteroscedasticity (ARCH) class of models, discrete time specifications within this class have been used extensively to model the so-called stylized features of financial returns. These properties, which include serially dependent volatility, heavy tails and asymmetry, are not captured by traditional linear time series like, say, the Autoregressive Moving Average (ARMA) model.  As an example, let $P_t$ denote the price of a financial stock or asset at time $t$, so that the time series of financial log-return is $\{Z_t := \log P_t-\log P_{t-1}\}$. Since the financial return $Z_t$ is essentially unpredictable in financial markets, $Z_t$ corresponds to the error term in a detrended ARMA specification, cf.\ Equations \ref{model} and \ref{eq:arma:model}, which is silent on the volatility dependence over time in $Z_t$, its heavy tails and its asymmetry. In finance, by contrast, since volatility  (``risk'') is predictable by the past, $Z_t$ is decomposed as
\begin{equation}\label{eq:Z:specification}
    Z_t = \sigma_t\varepsilon_t \,, \qquad \varepsilon_t\sim iid(0,1) \,, \qquad t\in \mathbb Z \,,
\end{equation}
where $\sigma_t$ is the conditional standard deviation or volatility (the predictable component) and an innovation $\varepsilon_t$ (the unpredictable component). The unpredictable $\varepsilon_t$'s can be heavy-tailed and asymmetric, and a key innovation in our paper is to use these features in the development of a flare detection method. Arguably, the most common volatility specification in finance is the Generalized ARCH (GARCH) model proposed by \citet{BOLLERSLEV1986}:
\begin{equation}\label{eq:garch:model}
    \sigma_t^2 = a_0 +
    \sum^p_{i=1}a_i Z^2_{t-i}+
    \sum^q_{j=1} b_j \sigma^2_{t-j} \,, 
\end{equation}
with $a_0 > 0$, $a_i \geq  0$, $i = 1,\dots, p$, and $b_j \geq  0$, $j = 1,\dots, q$ \citep[cf.][for a survey of GARCH models]{FrancqZakoian2019}. The structure of the model is similar to that of an ARMA model, and the model has been extensively used to predict future volatility by means of past values. In particular, the model has proved to be very useful in capturing the widespread volatility persistence feature of financial returns, namely that large (small) price fluctuations tend to be followed by more large (small) price fluctuations.  \cite{Stanislavsky2019} document that solar X-rays are characterized by the same heteroscedastic effects that are typically observed in finance in the form of volatility clustering, and demonstrate that the GARCH specification in Equation~\ref{eq:garch:model} can explain these properties. See Section~\ref{ARFIMA} for a comparison between our approach and the model adopted by \cite{Stanislavsky2019}. To illustrate, Figure \ref{fig:compar} illustrates how $Z_t$ and $\sigma_t$ of a financial return compare with those of a detrended light curve.
The variability patterns often analyzed in stellar observations share structural similarities with the financial data.  The fitted conditional volatility $\widehat{\sigma}_t$ of a GARCH(1,1) specification of DJUA log returns is shown as a function of time, capturing periods of heightened volatility and calmer intervals. The bottom right panel represents the fitted GARCH(1,1) conditional volatility $\widehat{\sigma}_t$ of the residuals $\widehat{Z}_t$ (see the analysis applied to stellar light curves below) of the detrended astronomical light curve, which exhibits similar volatility dynamics.

We describe the data we use in Section~\ref{sec:data}, the basis of our statistical model in Section~\ref{sec:model}, and the method we develop to detect flares in Section~\ref{sec:flaredetect}. We demonstrate its validity via simulations in Section~\ref{sec:simulation}, and apply it to exemplar TESS light curves in Section~\ref{sec:apply}.  We discuss and summarize our work in Sections~\ref{sec:discuss} and \ref{sec:summary}.

\section{Data: TESS Light Curves}\label{sec:data}

We demonstrate the application of our method on {selected} stars observed with TESS (Transiting Exoplanet Survey Satellite)\footnote{\url{https://tess.mit.edu}}.
{For our primary sample, we choose two stars} on which flares have been unambiguously observed \citep[see][]{Feinstein2022}, are at similar distances, but are otherwise unremarkable.  Neither are known to be exoplanet hosts.  \ticone\ is a dG5 type eruptive variable, and \tictwo\ is a low-mass, high proper motion dM4 type star.  
{In addition, we apply our method to the highly active flare star \aumic\ (\ticfour) as a point of validation and comparison (see Section~\ref{sec:compare} below).  \aumic\ is a well-studied, young ($\approx$20~Myr), low-mass ($\approx$0.6~M$_\odot$), nearby (9.7~pc) active dM1e dwarf with a luminosity $L_\textrm{bol}\approx0.1L_\odot$ and several confirmed exoplanets \citep[][]{2023MNRAS.525..455D}.  We estimate the luminosities of the less well-studied stars in our sample by scaling the observed TESS baseline fluxes to that of \aumic.  The basic properties and observation log of the program stars are listed in Table~\ref{tab:stars}.
}

We downloaded the processed light curve data from the TESS Science Processing Pipeline (TESS-SPOC)\footnote{\url{https://archive.stsci.edu/hlsp/tess-spoc}} from the MAST Portal (Mikulski Archive for Space Telescopes)\footnote{\url{https://archive.stsci.edu/missions-and-data/tess}} for the sectors listed in Table~\ref{tab:stars}.  The data are at several cadences.  We apply our method to detect flares in the {\tt PDCSAP} (Pre-Search Data Conditioning Simple Aperture Photometry) electron count rate light curves, which have backgrounds subtracted and systematics corrected for cotrending basis vectors\footnote{\url{https://spacetelescope.github.io/mast_notebooks/notebooks/TESS/beginner_tour_lc_tp/beginner_tour_lc_tp.html}}.  {We exclude times with no or bad data (marked with negative {\tt PDCSAP} flux values) in subsequent analysis, and treat each contiguous dataset individually.}

\begin{figure*}
    \centering
    \includegraphics[height=2.8in]{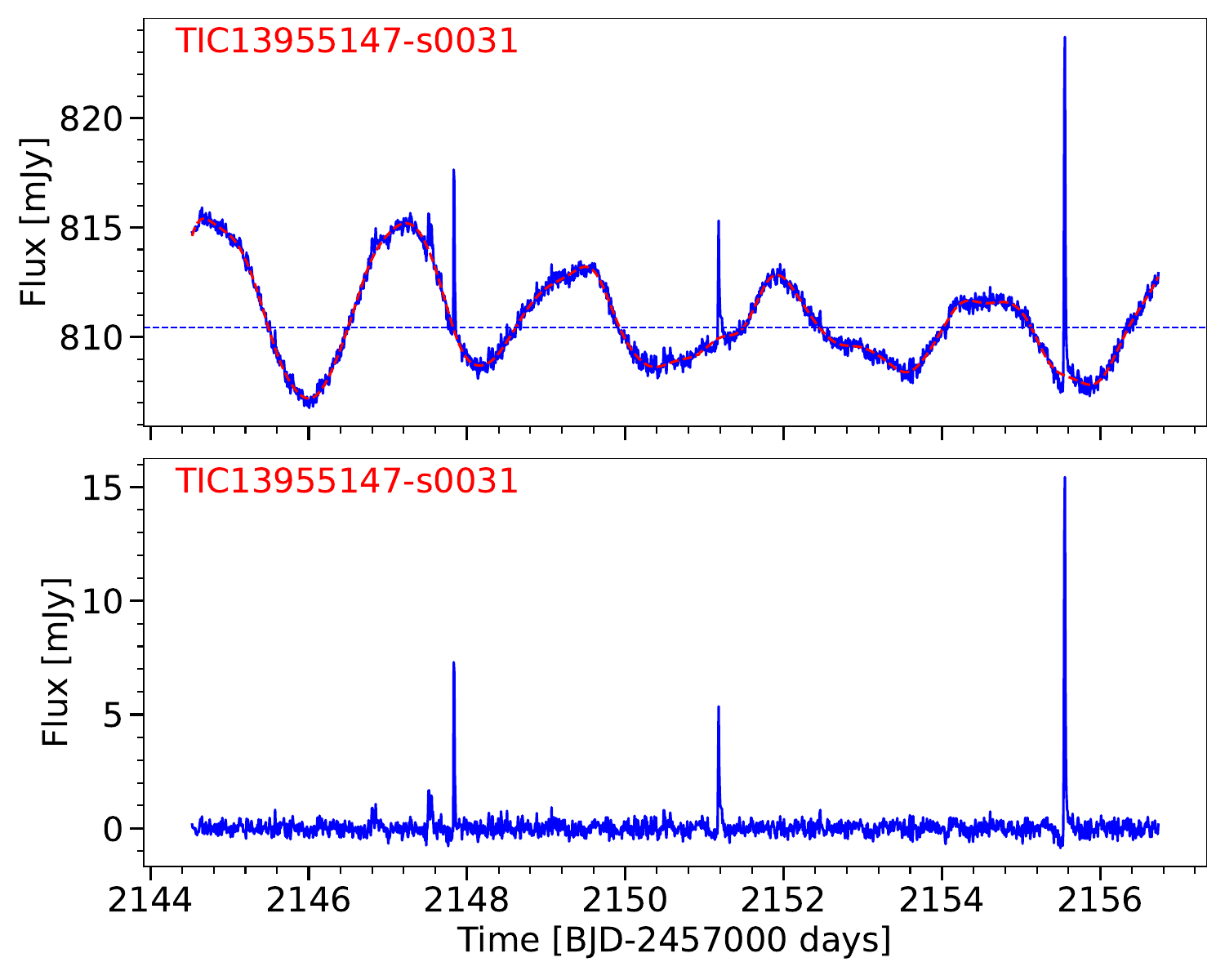}
    \includegraphics[height=2.8in]{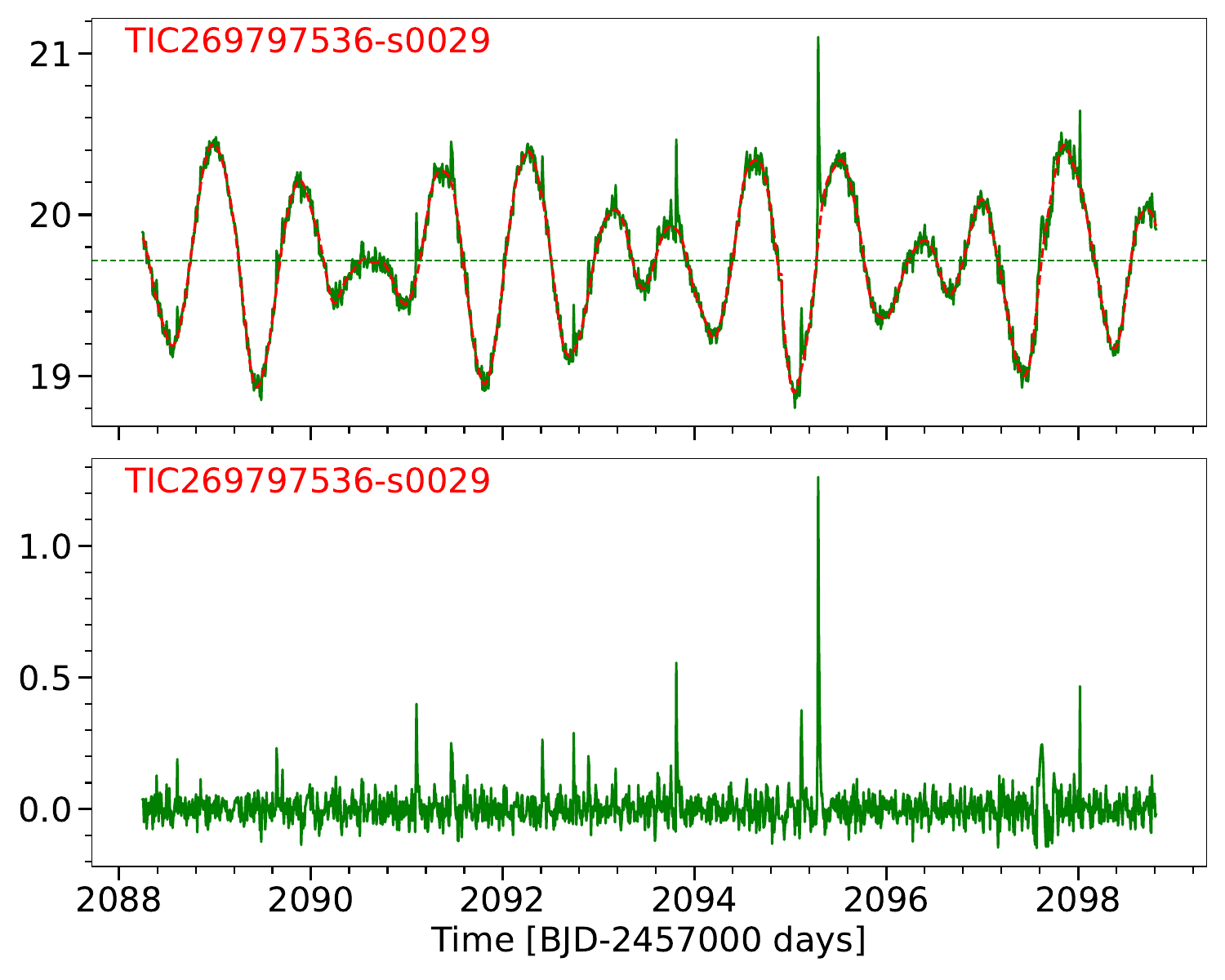}
    \caption{
    Illustrating the detrending of the baseline flux with a harmonic model.  {\sl Top panels:} The first halves of TESS light curves of \ticone\ sector 31 ({\sl left}) and \tictwo\ sector 29 ({\sl right}) are shown, with the harmonic detrended curve overplotted as a dashed curve.  {\sl Bottom panels:} The residuals after harmonic detrending are shown for the corresponding datasets.
    }
    \label{fig:harmonic_fitting}
\end{figure*}

We show exemplar light curves in the upper panels of Figure~\ref{fig:harmonic_fitting}, for the first halves of the data from \ticone\ Sector 31 and \tictwo\ Sector 29.  Note that the intensities are shown converted from electron count rate to mJy\footnote{The measured count rates $r$ in [e$^{-}$~s$^{-1}$] are first converted to TESS magnitudes $m_T = -2.5\log_{10}(r)+20.44$, and then to fluxes as $2.416 \times 10^{6} \times 10^{-0.4{\cdot}m_T}$ [mJy] (see \url{https://tess.mit.edu/public/tesstransients/pages/readme.html}).  
{Further, scaling the baseline flux $4567.35$~mJy of \aumic\ to its nominal $L_\textrm{bol}=0.1 L_\odot$, we convert $1~\mathrm{mJy}\Leftrightarrow{7.42}\times{10^{-12}}$~ergs~s$^{-1}$~cm$^{-2}$.}
%Further, adopting a passband width of $2\times10^{14}$~Hz based on the TESS bandpass of 600-1000~$\mu$m, we also convert $1~\mathrm{mJy}:\Rightarrow{2}\times{10^{-12}}$~ergs~s$^{-1}$~cm$^{-2}$.
\label{foot:rate2fx}}.  We illustrate our method by showing the application of each step of our method on these light curves\footnote{Corresponding figures for {\sl all} of the data sectors are available in a Zenodo repository \citep[\url{https://doi.org/10.5281/zenodo.14991246};][]{wang_2025_14991246}}.  {Similar figures for the Sector 1 data of \ticfour\ (\aumic) are shown in Appendix~\ref{app:aumic}.}

\begin{table*}
    \centering
    \caption{{List of stars and datasets selected}}
    \begin{tabular}{lcccccl}
    \hline\hline
    TIC & Other Names$^\dag$ & Spectral & Distance$^\dag$ & {Baseline flux}$^*$ & Cadence$^\ddag$ & Sectors \\
    \hfil & \hfil & Type$^\dag$ & [pc] & {[mJy]} & [min] & \hfil \\
    \hline
    {\href{https://mast.stsci.edu/portal/Mashup/Clients/Mast/Portal.html?searchQuery=%7B%22service%22%3A%22CAOMDB%22%2C%22inputText%22%3A%22TIC%2013955147%22%2C%22paramsService%22%3A%22Mast.Caom.Cone%22%2C%22title%22%3A%22MAST%3A%20TIC%2013955147%22%2C%22columns%22%3A%22*%22%2C%22caomVersion%22%3Anull%7D}{13955147}} & HD\,32372 & G5\,V & 78 & $830 \pm 23.5$ & 30 & 4, 5 \\ 
    \hfil & 2MASS\,J05005186-4101065 & \hfil & \hfil & \hfil & 10 & 31, 32 \\
    \hfil & Gaia\,DR3\,4813691219557127808 & \multicolumn{5}{c}{\hfil} \\ 
    \hfil & 1RXS\,J050051.7-410100 & \multicolumn{5}{c}{\hfil} \\ 
    \hline
    {\href{https://mast.stsci.edu/portal/Mashup/Clients/Mast/Portal.html?searchQuery=%7B%22service%22%3A%22CAOMDB%22%2C%22inputText%22%3A%22TIC%20269797536%22%2C%22paramsService%22%3A%22Mast.Caom.Cone%22%2C%22title%22%3A%22MAST%3A%20TIC%20269797536%22%2C%22columns%22%3A%22*%22%2C%22caomVersion%22%3Anull%7D}{269797536}} & 2MASS\,J04363294-7851021 & M4\,V & 70 & $19.7 \pm 0.4$ & 30 & 2, 5, 8, 11, 12, 13,  \\
    \hfil & Gaia\,DR3\,4622912654918835200 & \hfil & \hfil & \hfil & 10 & 27, 28, 29, 32, 35, 38, 39 \\
    \hline
    {\href{https://mast.stsci.edu/portal/Mashup/Clients/Mast/Portal.html?searchQuery=%7B%22service%22%3A%22CAOMDB%22%2C%22inputText%22%3A%22TIC%20441420236%22%2C%22paramsService%22%3A%22Mast.Caom.Cone%22%2C%22title%22%3A%22MAST%3A%20TIC%20441420236%22%2C%22columns%22%3A%22*%22%2C%22caomVersion%22%3Anull%7D}{441420236}} & AU\,Mic, HD\,197481 & M1Ve & 9.71 & $4567 \pm 80$ & 2 & 1, 27 \\
    \hfil & 2MASS\,J20450949-3120266 & \multicolumn{5}{l}{\hfil} \\
    \hfil & Gaia\,DR3\,6794047652729201024 & \multicolumn{5}{l}{\hfil} \\
    \hline
    \multicolumn{6}{l}{$\dag$: From the SIMBAD  database \citep{2000A&AS..143....9W}.} \\
    \multicolumn{6}{l}{{$*$: Excluding candidate flares (see text), and averaged over all sectors.}} \\
    \multicolumn{6}{l}{$\ddag$: The exposure frame time is 2~s; the number of frames used to create the processed light curve varies (60, 300, or 600).} \\
    \end{tabular}
    \label{tab:stars}
\end{table*}

\section{Our statistical model}
\label{sec:model}

We describe the light curves using a novel model where the observed data $Y_t$ is composed of a deterministic trend $\mu(t)$ and a stochastic component $X_t$,
\begin{equation}\label{model}
Y_t= \mu(t) + X_t,\qquad t=1,\dots,n,
\end{equation}
where $n$ is the sample size. The function $\mu(t)$ is a deterministic, periodic time-trend that allows for time-variation in its coefficients \citep[as in][]{MSC2022}, whereas $X_t$ follows a stationary ARMA($r,s$) model,
\begin{equation}
    X_t = \sum_{i=1}^r \phi_i X_{t-i} + \sum_{j=1}^s \theta_j Z_{t-j} + Z_t \,, \label{eq:arma:model}
\end{equation}
with $Z_t$ governed by the GARCH($p,q$) model given by Equation~\ref{eq:Z:specification} and Equation~\ref{eq:garch:model}.
Accordingly, the conditional prediction of $Y_t$ is
$\mu(t) + \sum_{i=1}^r \phi_i X_{t-i} + \sum_{j=1}^s \theta_j Z_{t-j}\,,$ 
and the conditional prediction of the variability $Z_t^2$ is given by $\sigma_t^2$ as in Equation~\ref{eq:Z:specification}. Henceforth, we focus on the GARCH specification of $Z_t$ in Equation~\ref{eq:arma:model} since we use this structure to develop our new flare detection method.

As is evident in the light curves (see Figure~\ref{fig:harmonic_fitting}), the baseline shows distinct periodic variations with varying amplitudes, at a timescale distinct from that of flares.  Since the baseline emission is not blocked by opacity effects during flares, it is reasonable to smoothly interpolate across flaring intervals to determine the baseline trends.  We thus adopt a two-stage process where we first detrend the light curve to remove the harmonic variations, and analyze the residuals to then detect flare events.  
{Several methods have been used in the literature to determine long time scale trends, ranging from using so-called sigma-clipping \citep[see description and critique in][]{2005ApJ...629..172N}, split wavelet bases at appropriate scales \citep[][]{10.1007/978-1-4614-3520-4_16,2013arXiv1301.3027B}, fitting splines \citep[to supernovae light curves, by][]{Dhawan2015}, using flexible ARIMA models \citep[to Kepler light curves, by][]{F2018}, and specifically to TESS light curves, using iterative Savitzsky-Golay filters fitted with 4$^{th}$-order polynomials \citep[][]{2022AJ....163..147G}, period finding and median filtering at two time scales \citep[][]{2020AJ....159...60G,2020AJ....160..219F}, and various forms of Gaussian Process regression \citep[][]{2022A&A...661A.148C,2025ApJ...979..141E,2025arXiv250515451G}.}
These forms are highly flexible, but for that same reason also tend to overfit the baseline as they do not explicitly take into account the intensity and period changes that occur due to rotational modulations of intermittent differential spottedness of the stellar photospheres.  Thus, they are likely to introduce biases in both detectabilty and flux estimation.  Indeed, \cite{2025ApJ...979..141E} explicitly caution against using techniques like sigma-clipping to locate flares.  

{Here, we adopt a more physically motivated model that can account for starspot occurrence variations by using a linear combinations of sinusoidal waves with time-varying amplitudes, a simplified version of the model introduced by \cite{MSC2022}.}
The inclusion of a time varying amplitude makes the model sufficiently flexible to capture the periodic changes in amplitude.  Our model takes the form
\begin{equation}\label{HF}
\mu(t)= \omega_0(t)+\sum_{k=1}^K \omega_k(t) \sin \left(2 \pi \frac{t}{\tau} k + \eta_k\right),
\end{equation}
where $t$ are the observation times with $t=0$ on day 2457000~BJD, and $\omega_k(t)$ are time-varying amplitudes,
\begin{eqnarray}
\omega_0(t) &=& \gamma_0 \beta_0+\beta_0 \sin \left(2 \pi \frac{t}{\tau}+\varphi_0\right) \, \nonumber \\
\omega_k(t) &=& \gamma_0 \gamma_k+\gamma_k \sin \left(2 \pi \frac{t}{\tau} k +\varphi_k\right) \,,
\label{eqn:harmonic_amplitudes}
\end{eqnarray}
with $\tau$, $\gamma_k$, $\beta_0$, $\eta_k$, $\varphi_k$ being the fitted hyperparameters, and $\eta_k,\varphi_k$ (in radians) in particular being phase shifts relative to $t=\textrm{BJD}~2457000$.  We typically use $K=20$ terms in the model.

\section{Flare Detection}\label{sec:flaredetect}

Our algorithm has two parts. In the first part (see section \ref{NPD}, as well as the first ten lines of Algorithm~\ref{algo} below) we iteratively remove the harmonic trend from the observed light curve. {If uncorrected,} the fit of the harmonic trend is sensitive to the presence of flares{, because the strong, short-lived flux enhancements associated with flares bias the estimation of the time-varying amplitudes $\omega_k(t)$ and $\eta_k$ in Eq.~\eqref{HF}–\eqref{eqn:harmonic_amplitudes}. When included in the fit, these outliers can distort the periodic modulation by artificially increasing the fitted amplitudes or shifting the phases. To prevent this leakage of flare power into the harmonic component, our algorithm iteratively identifies and removes flare-affected points before re-estimating $\omega_k(t)$}. We therefore remove, at each iteration, the flares from the light curve  as well as the time-varying trend (see points 5 and 8 in Algorithm~\ref{algo} below).
{In the second part (see Section~\ref{PFD} below), we detect flares as significant positive deviants of the standardized model residuals $\varepsilon_t$ (see lines 11 and above of Algorithm~\ref{algo}).}

\begin{algorithm}
\caption{Flare-detection and model-update}\label{algo}
\begin{algorithmic}[1]
%Non-parametric detrending
\State Define $\boldsymbol X^{(0)}=\boldsymbol  Y^{(0)} 
- \boldsymbol \mu^{(0)}$ and $\alpha^{(0)}=\alpha_{\rm max}$
\For{$i$ in $0:(I-1)$}
\State Fit an ARMA($r,s)$ to $ \boldsymbol  X^{(i)}$ so to obtain $\boldsymbol 
 Z^{(i)}=\tfrac{\phi(B)^{(i)}}{\theta(B)^{(i)}}\boldsymbol 
 X^{(i)}$ 
\ForAll{$t$ in 1:n}\,\, $\pi_{t}^{(i)}={\rm Pr} \big(\mathcal{H}>\frac{|Z_t^{(i)}|}{S_t^{(i)}}\big)$
\If{$\pi_{t}^{(i)}<\alpha^{(i)}$}
%{$\frac{X_t^{(i)}}{S_t^{(i)}} > Q^{HN}_{(\alpha)}$}
\State delete $\{t:t+9, Y_t:Y_{t+9}\}$ %\qquad\mbox{(20 minutes forwards)}
\Else
\State{$Y_t^{(i+1)}=Y^{(i)}_t$}   
\EndIf
\EndFor
\State fit $\boldsymbol \mu^{(i+1)}$
to $\boldsymbol  Y^{(i+1)}$
\State  $\boldsymbol  X^{(i+1)}=\boldsymbol  Y^{(i+1)} - \boldsymbol \mu^{(i+1)}$, and  $\alpha^{(i+1)}
=\displaystyle\max_{t:\,\pi_{t}^{(i)}<\alpha^{(i)}}
\pi_{t}^{(i)}$
\EndFor
\State // Fit an ARMA($r,s$)-GARCH($p,q$) to $\boldsymbol  Y^{(0)}- \boldsymbol  \mu^{(I)}$
{
\State Initialize best BIC score $\mathcal{B}_{\text{best}} \gets \infty$
\ForAll{ARMA orders $(r,s) \in \{1,3\}$}
    \ForAll{GARCH orders $(p,q) \in \{1,3\}$}
        \State Specify model: conditional mean $\mu_t$ as ARMA($r,s$), conditional variance $\sigma_t^2$ as GARCH($p,q$)
        \State Estimate model parameters jointly using maximum likelihood (MLE)
        \State Compute BIC $\mathcal{B}_{r,s,p,q}$
        \If{$\mathcal{B}_{r,s,p,q} < \mathcal{B}_{\text{best}}$}
            \State Update best model and set $\mathcal{B}_{\text{best}} \gets \mathcal{B}_{r,s,p,q}$
        \EndIf
    \EndFor
\EndFor
\State Retain standardized residuals $\boldsymbol \varepsilon_t = \boldsymbol Z_t / \boldsymbol \sigma_t$ from best model
}

\ForAll{$t$ in 1:n}\,\, $q_{(1-\alpha)}^{0} = \Phi^{-1}(1-\alpha)$
\If{$\varepsilon_{t}>q_{(1-\alpha)}^{0}$}
%{$\frac{X_t^{(i)}}{S_t^{(i)}} > Q^{HN}_{(\alpha)}$}
Define {flare candidates} $F_t= Y_t$
%\Else
%\State{$Y_t^{(i+1)}=Y^{(i)}_t$} 
\EndIf
\EndFor
\end{algorithmic}
\end{algorithm}

\begin{figure*}
\centering
    \includegraphics[height=2.8in]{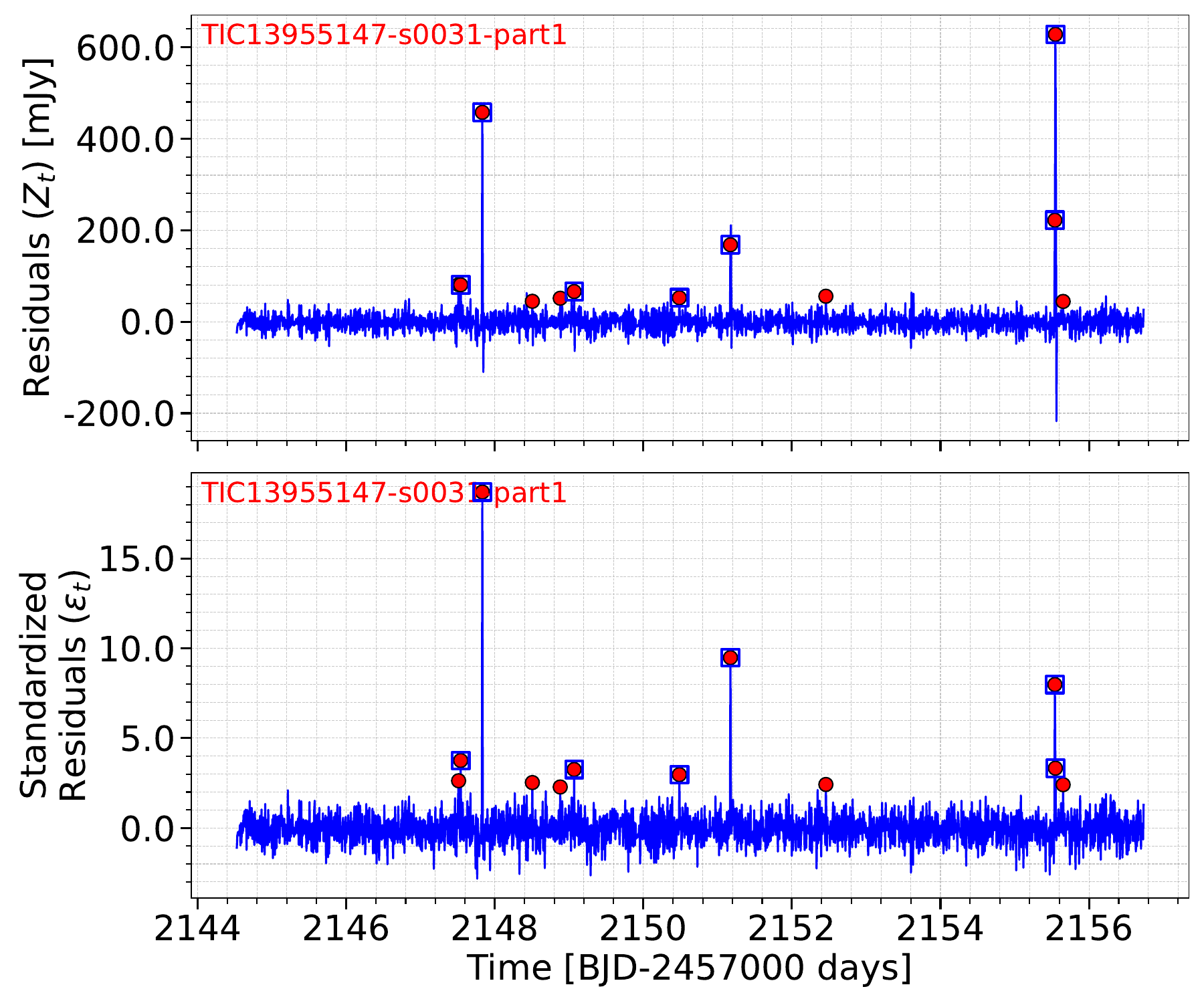}
    \includegraphics[height=2.8in]{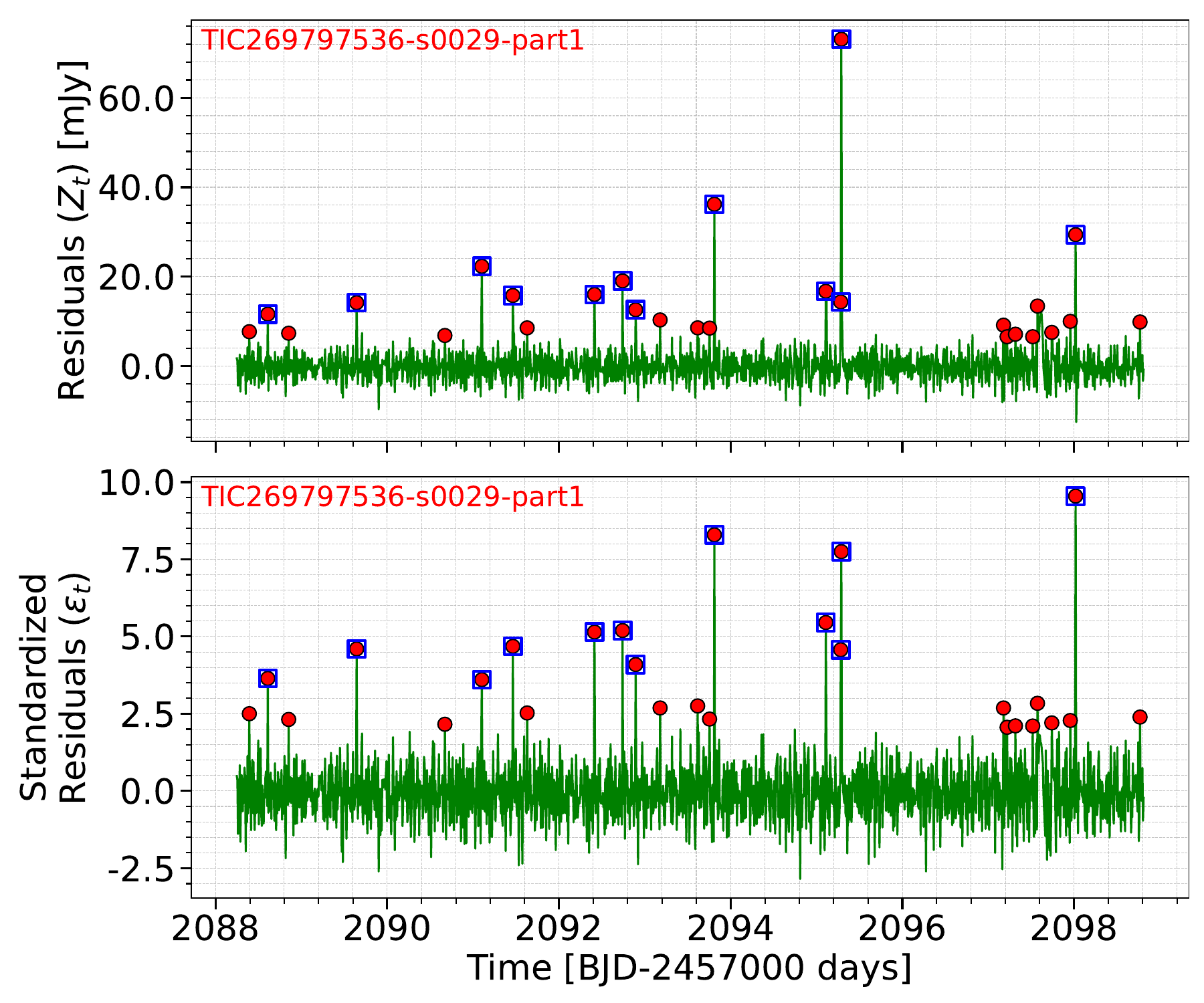}
    \caption{
    Illustrating the analysis of the residuals after harmonic detrending for the light curves in Figure~\ref{fig:compar}, with the detected flares passed BH procedure marked as solid dots and passed HB procedure as hover squares.  {\sl Top panels:} Residuals $Z_t$ after detrending with the harmonic fit {and bypassing the ARMA-GARCH filter}.  {\sl Bottom panels:} Standardized residuals $\hat\varepsilon_t=\hat{Z}_t / \hat{\sigma}_t$, where $\hat{Z}_t$ is the ARMA($r,s$) residual and $\hat{\sigma}_t$ is the estimated conditional volatility.
    }
\label{fig:quatile_detect_demo}
\end{figure*}

\begin{figure*}
\centering
    \includegraphics[height=2.8in]{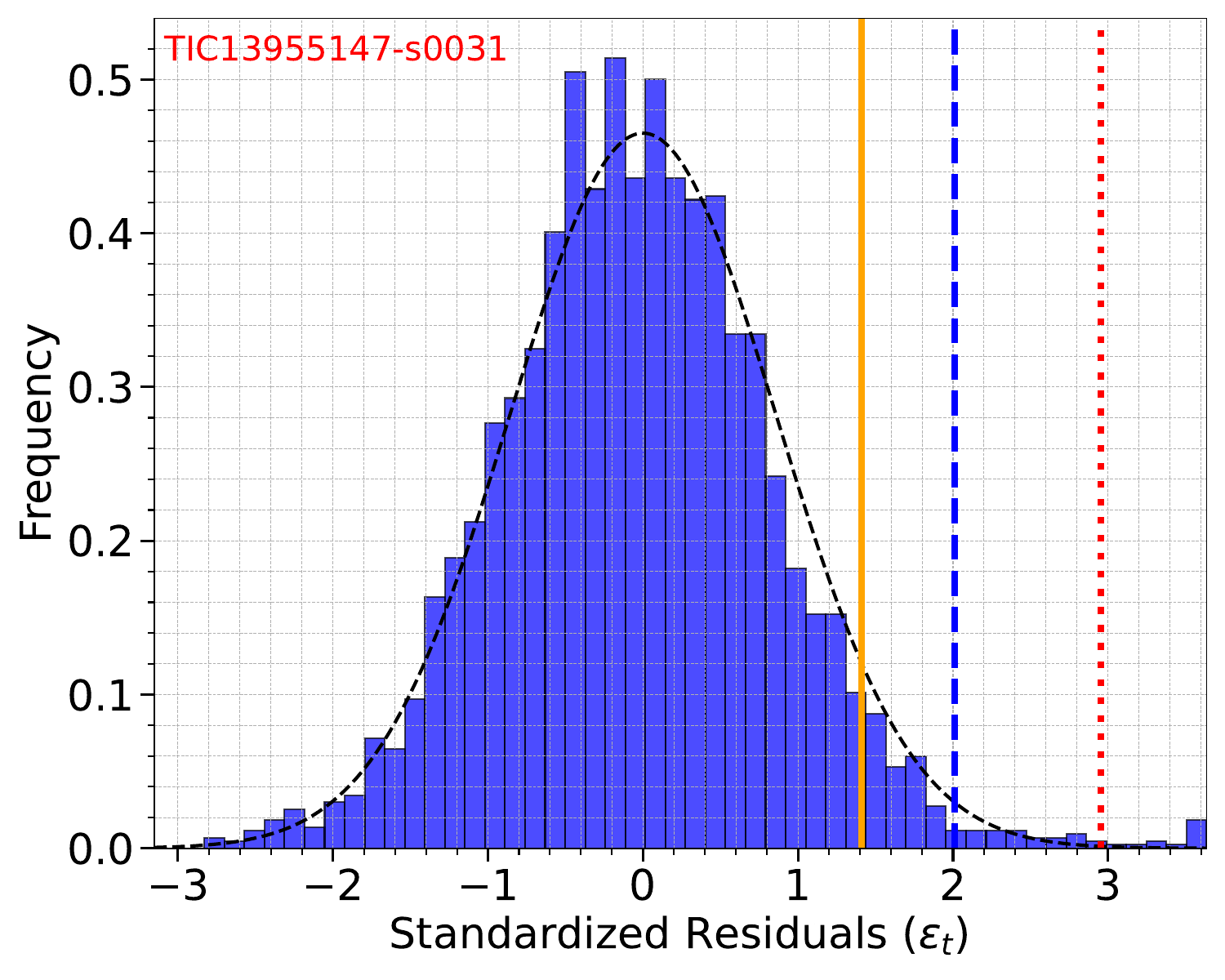}
    \includegraphics[height=2.8in]{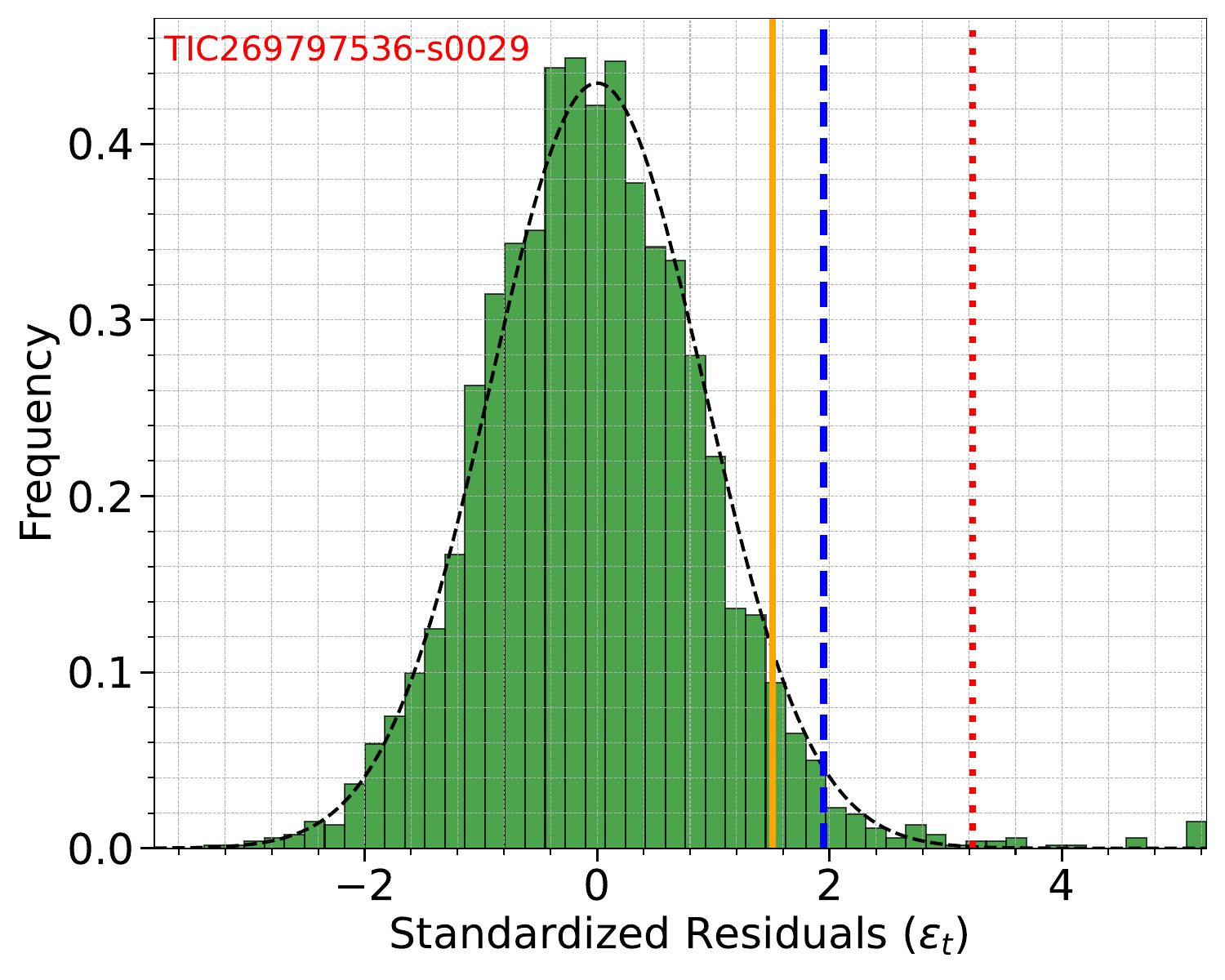}
    \caption{
    Distributions of the standardized residuals $\varepsilon_t$ for the data sets shown in Figure~\ref{fig:quatile_detect_demo}.  The vertical solid line marks the empirical 95\% upper quantile of $\varepsilon_t$, the line dashed vertical line marks the BH threshold under $\alpha = 0.05$ while the dot dashed line marks the HB threshold, and the dashed curve is the fitted normal distribution shown for comparison.  Time bins which exceed the 95\% quantile are chosen as candidate flares.
    }
\label{fig:epsilont_dist}
\end{figure*}

\subsection{Semi-parametric detrending}\label{NPD}

The first challenge we face is the removal of the time-varying trend $\mu(t)$. To this end, we fit the model in Equation~\ref{HF} to the observed light curve $Y_t$. In order to fit the trend accurately, the pairs $\{t,Y_t\}$ corresponding to the flares must be excluded from the fit.
We detail the procedure in lines 1-10 of Algorithm~\ref{algo}.  
%Briefly, we use the principles of hypothesis testing to find and 
{Briefly, we filter on the corresponding $p$-values $\pi_t$ to} iteratively eliminate intervals likely to contain flares, and fit a flexible variable amplitude harmonic model (see Equations~\ref{HF}-\ref{eqn:harmonic_amplitudes}) to the remaining data.  

Our model in 
Equation~\ref{model}-\ref{eq:arma:model} can be written as 
\begin{equation}
\begin{split}
Y_t &=\mu(t) +X_t \,,\\
\phi(B)X_t&=\theta(B)Z_t \,,\\
Z_t&=\sigma_t \varepsilon_t \,, \\ %\qquad
\varepsilon_t &\sim iid(0,1) \,,
\label{eqn:arma}
\end{split}
\end{equation}
where $B$ is the back-shift operator\footnote{The operator $B$  shifts time-indexing by $-1$, i.e., $By_t:=y_{t-1}$.  Multiple applications of $B$ can be represented algebraically, e.g., $B(B\,y_t):=B^2\,y_t{\equiv}y_{t-2}$, first order differencing becomes $(1-B)y_t:=y_t - y_{t-1}$ and second order differencing becomes $(1-B)^2y_t:=(1-2B+B^2)\,y_t{\equiv}y_t-2y_{t-1}+y_{t-2}$, etc.  It is often convenient to use this within complex forms as in Equation~\eqref{eqn:arma}.}, and the polynomial filters $\phi(B)=1-\sum_{j=1}^r\phi_jB^j$ and $\theta(B)=1+\sum_{k=1}^s\theta_jB^k$ do not have roots inside the unit circle (these conditions ensure that the ARMA process is causal and invertible).  In the first step, we identify flares in the TESS light curves via hypothesis testing.  We {use} a uniform  threshold, computed using the residuals of a light curve, that is $t$-invariant and $n$-invariant, yet examining each time-varying volatility within its local neighborhood. Given the residual value 
\begin{equation}
Z_t=\tfrac{\phi(B)}{\theta(B)}[Y_t-\mu(t)] 
\end{equation}
obtained by removing the fitted trend $\mu(t)$ from the observed light curve $Y_t$ and by fitting the ARMA model in Equation~\ref{eq:arma:model} to $X_t=Y_t -\mu(t)$, we adopt a non-parametric test statistic that is easy to compute. Since the conditional variance of a GARCH is time-varying, we can test the null hypothesis of  $Z_t|\mathcal I_t$ being a white noise with $\{\operatorname{Var}(Z_t|\mathcal I_t)=\sigma^2_0$ for all $t\}$ versus the alternative $\{\operatorname{Var}(Z_t|\mathcal I_t)=\sigma^2_t>\sigma^2_0\,\mbox{for some}\, t\}$. To this end, we define our time-varying test statistic as ${Z_t^2}/{S_t^2}$, where $S_t$ is a non-parametric estimator of the conditional volatility $\sigma_t$. We estimate $\sigma_t$ non-parametrically because our harmonic trend-function $\mu(t)$ is time-varying and needs to be fitted locally in time rather than globally. Under the null $Z_t|\mathcal{I}_t=\sigma_0\varepsilon_t$ is Gaussian with mean zero and variance $\sigma^2_0$ for all $t$, whereas under the alternative $Z_t|\mathcal I_t$ is Gaussian with mean zero and variance $\sigma^2_t>\sigma^2_0$. Under the null, ${Z_t^2}/{S_t^2}$ is distributed according to a $F_{1,n}$-distribution, where $n$ is the sample size. For a given $\alpha$ at time $t$ we reject the null if ${Z_t^2}/{S_t^2}>F_{1,n-1}(1-\alpha)$.

The ratio ${Z_t}/{S_t}$ is a $t$-student distribution with $n-1$ degrees of freedom, and its square follows the $F$-distribution with degrees of freedom 1 and $n-1$.  As $m \rightarrow \infty$ the random variable $F_{\ell,m}$ converges in distribution to a $\chi^2_\ell$, the chi-square distribution with $\ell$ degrees of freedom \citep{casella_berger_2002}.  Since in our test the parameter $\ell$ is fixed and equal to one, this results provides a mathematical argument to adopt a uniform threshold given by $\chi^2_{\!_1(\alpha)}$, the $\alpha$-quantile of the $\chi^2_1$ distribution. This quantile is very close (for large $n$) to the finite-sample $F_{1,n-1}(\alpha)$ threshold, as long as the sample size $n$ is large enough. The threshold is `uniform' in the sense that it does not depend on either $t$ or $n$. 

Next notice that the condition ${\rm sign} \{Z_t\}\times \big({Z_t}/{S_t}\big)^2>\chi^2_{\!_1(\alpha)}$ is equivalent to ${Z_t}/{S_t} > Q^{\scalebox{0.75}{HN}}_{(\alpha)}$, where $\alpha={\rm Pr} \big(\mathcal{H}>Q^{\scalebox{0.75}{HN}}_{(\alpha)}\big)$, $\mathcal{H}$ being a Standard Half-Normal random variable. Hence, the definition of our rejection region can be simplified as $\{\,{Z_t}> {S_t}\,Q^{\scalebox{0.75}{HN}}_{(\alpha)}\}$. This definition discloses an intuitive interpretation of our approach. The flares are values of (the luminosity of) the light curve significantly exceeding the periodic trend. They can be defined as those pairs $\{t,Y_t\}$ such that $Z_t> Q \sigma_0$, and our testing procedure is thus `naturally mimicking' the presence of a flare.  In practice we suggest to compute $S_t$ using a robust method, such as the estimator suggested by \cite{RC1993}. Their estimator is an improved version of the Median Absolute Deviation, which is limited to symmetric distributions and has low efficiency at Gaussian distributions.

We reject the null if ${\rm Pr} \big(\mathcal{H}>Q^{\scalebox{0.75}{HN}}_{(\alpha)}\big)<\alpha$, where the tuning parameter $\alpha$ (the significance level of the test) is often set {\it ex-ante}, that is, {\it before} performing the test. We might instead be interested in determining $\alpha$ in a data-driven way. In other words, we could aim at estimating $\alpha$ rather than assigning an arbitrary value to this parameter. This goal can be achieved by defining an upper bound $\alpha_{\rm max}$ and then selecting, among those values of the above probability that are smaller than $\alpha_{\rm max}$, the largest one.

Since the harmonic fit $\mu$ is computed from the observed data $Y$, the presence of flares which span multiple bins makes the estimated $\mu$  biased. For this reason, for every time bin with a suspected flare, we also exclude an additional nine succeeding bins (corresponding to durations of $\approx$5-15~ks) and then re-fit $\mu$ to the `flares-filtered' data.

All the arguments of the above discussion lead to lines 1-10 of Algorithm~\ref{algo}. In it, $Y^{(0)}$ is the observed time series and $\mu^{(i)}$ is the harmonic trend in Equation~\ref{HF} fitted to $Y^{(i)}$, $0\leq i \leq  I-1$, where $I$ is the number of iterations. The value of $\alpha$ is guaranteed to decrease over $i$ because the inequality appearing in both the condition $\pi_{t}^{(i)}<\alpha^{(i)}$ and the update $\alpha^{(i+1)}$ is strict.

\subsection{Parametric ARMA-GARCH flares detection}\label{PFD}

Here we describe the second stage of our method, to detect candidate flares in the detrended residual light curves, listed in {lines 11-22} of Algorithm~\ref{algo}.  
{Our method relies on the property of ARMA and GARCH to capture the correlation structure present in the detrended residuals $X_t$, yielding a model of residuals $Z_t$ with GARCH conditional variances $\sigma_t$.  At this stage, in the absence of any signal, the $\varepsilon_t$ should be distributed nominally as uncorrelated values drawn from a Gaussian distribution with mean zero.  Note that the lack of a ``signal'' does not mean that the light curve is flat or is characterized only by statistical fluctuations; there may yet exist autocorrelation structure akin to emission from a quiescent corona.  Any instance that shows a departure from the nominal behavior of the star, such as the sudden appearance of impulsive deviations like flares, will then be captured as deviations in the distribution of $\varepsilon_t$.}
The candidate flares found by this process are analyzed further to balance false positives against false negatives in Section~\ref{sec:HB_BH}.  {In the following, we describe the mathematics of the procedure in detail.}

%Let $E_{t-1}(X_t) := \sum_{i=1}^r \phi_i X_{t-i} + \sum_{j=1}^s \theta_j Z_{t-j}$ denote the ARMA prediction at $t$ conditional on the past, i.e.\ values of $X_t$ and $Z_t$ up to and including $t-1$ (hence the subscript $t-1$).  
{Let $\mu_X(t) := \sum_{i=1}^r \phi_i X_{t-i} + \sum_{j=1}^s \theta_j Z_{t-j}$ denote the ARMA prediction of $X_t$. Consequently, Equation~\ref{eq:arma:model} can be re-written as $X_t = \mu_X(t) + Z_t$ with $Z_t=\sigma_t\varepsilon_t$, so that $Y_t = \mu(t) + \mu_X(t) + \sigma_t\varepsilon_t$.} The orders $r$ and $s$ of the ARMA model for $X_t$, as well as the {orders of} the GARCH model for $Z_t$, are selected using BIC (Bayesian Information Criterion). The parameters of the ARMA {model}, as well as those of the GARCH {model}, are estimated jointly by Maximum Likelihood {as} discussed by \cite{FrancqZakoian2004}. {For estimation we use the \textsf{R} package \texttt{rugarch} \citep{Ghalanos2024rugarch}, which provides a flexible and efficient framework for fitting ARMA-GARCH-type models. Model selection is conducted over a grid with ARMA and GARCH orders up from 1 up to 3.}

%The orders $r$ and $s$ of the ARMA model for $X_t$, as well as the orders $p$ and $q$ of the GARCH model for $Z_t$, are selected using BIC (Bayesian Information Criterion). The parameters of the ARMA filter, as well as those of the GARCH filter, are estimated jointly by Maximum Likelihood discussed by \cite{FrancqZakoian2004}.  {For this estimation, we use the \texttt{rugarch} package in \textsf{R} created by \cite{Ghalanos2024rugarch}, which provides a flexible and efficient framework for fitting ARMA-GARCH-type models. The model selection is conducted over a grid with ARMA and GARCH orders up from 1 up to 3. Consequently,} Equation~\ref{eq:arma:model} can be re-written as $X_t = E_{t-1}(X_t) + Z_t$ with $Z_t=\sigma_t\varepsilon_t$, so that $Y_t = \mu(t) + E_{t-1}(X_t) + \sigma_t\varepsilon_t$. 

The {conditional} probability distribution of $Y_t$ {is obtained as a} straightforward transformation of the probability distribution of $\varepsilon_t$. {The conditional cumulative distribution of $\varepsilon_t$ is $F(\varepsilon_t) = Pr(\varepsilon_t \leq q)$, which is constant over time due to the the \textit{iid} property of $\varepsilon_t$ (cf.\ Equation~\ref{eq:Z:specification}). The conditional cumulative distribution of $Y_t$, by contrast, is time-varying and equal to
\begin{equation*}
	Pr(\varepsilon_t \leq q) \, = \, Pr\bigr(Y_t \leq y(t)\bigr) \quad\text{with}\quad y(t) = \mu(t) + \mu_X(t) + \sigma_t q.
\end{equation*}
This follows from $\varepsilon_t = \bigr(Y_t - \mu(t) - \mu_X(t)\bigr)/\sigma_t$. While the conditional $(1-\alpha)$-quantile of $\varepsilon_t$ is constant over time and equal to $q_{(1-\alpha)} = F^{-1}(1-\alpha)$, the conditional $(1-\alpha)$-quantile of $Y_t$ is time-varying and equal to $y_{(1-\alpha)}(t) = \mu(t) + {\mu_X(t)} + \sigma_t q_{(1-\alpha)}$, since
\begin{equation*}
	Pr(\varepsilon_t \leq q_{(1-\alpha)}) \; = \; Pr\bigr( Y_t \leq \mu(t) + \mu_X(t) + \sigma_t q_{(1-\alpha)} \bigr).
\end{equation*}
Again, this follows from $\varepsilon_t = \bigr(Y_t - \mu(t) - \mu_X(t)\bigr)/\sigma_t$.}

%Specifically, the $(1-\alpha)$-quantile of $\varepsilon_t$ conditional on the past is $q_{(1-\alpha)} = F^{-1}(1-\alpha)$, which is constant over time due to Equation~\ref{eq:Z:specification}, where $F$ is the density of $\varepsilon_t$. By contrast, the $(1-\alpha)$-quantile of $Y_t$ conditional on the past, which is time-varying, is $y_{(1-\alpha)}(t) = \mu(t) + E_{t-1}(X_t) + \sigma_t q_{(1-\alpha)}$.

Our method for detecting {candidate} stellar flares consists of classifying observations $Y_t$'s as stellar flares if they exceed the time-varying conditional $(1-\alpha)$-quantile derived under the assumption that no flare occur. In the expression for $y_{(1-\alpha)}(t)$, this amounts to replacing $q_{(1-\alpha)}$ with a quantile derived under the assumption that no flare occur. Henceforth, we denote this ``no flare'' conditional quantile for $q_{(1-\alpha)}^{0}$, so that the time-varying threshold becomes
\begin{equation}\label{threshold}
    y_{(1-\alpha)}^{0}(t) = \mu(t) + {\mu_X(t)} + \sigma_t q_{(1-\alpha)}^{0}.
\end{equation}
To understand the origin of $q_{(1-\alpha)}^{0}$, note that the variance of $\varepsilon_t$ can be written as $E(\varepsilon_t^2) = Pr(\varepsilon_t {\leq} 0) E(\varepsilon_t^2 | \varepsilon_t {\leq} 0) + Pr(\varepsilon_t {>} 0) E(\varepsilon_t^2 | \varepsilon_t {>} 0)$. Let $A_t$ denote the event that a flare occurs at $t$ and let $A_t^C$ denote the event that a flare does not occur at $t$. The occurrence of flares is associated with {positive} $\varepsilon_t$'s. Accordingly, under the assumption that $\varepsilon_t$ has the same variance for both negative and positive values when no flare occurs at $t$, we have
\begin{equation}
    E(\varepsilon_t^2 | \varepsilon_t {\leq} 0) = E(\varepsilon_t^2 | \varepsilon_t {\leq} 0 \cap A_t^C) = E(\varepsilon_t^2 | \varepsilon_t {>} 0 \cap A_t^C) \,.
\end{equation}

{The distribution of $\varepsilon_t$ conditional on no flare occurring at $t$ is $F(\varepsilon_t | A_t^C)$. To make our procedure operational, we set}
\begin{equation}
    \Phi(\varepsilon_t) := N\bigr(0, E(\varepsilon_t^2 | \varepsilon_t {\leq} 0) \bigr) = F(\varepsilon_t | A_t^C)
\end{equation}
%denote the conditional distribution under no flare that we use in obtaining $q_{(1-\alpha)}^{0}$, so that $q_{(1-\alpha)}^{0} := \Phi^{-1}(1-\alpha)$. 
{so that $q_{(1-\alpha)}^{0} := \Phi^{-1}(1-\alpha)$}. {In other words, we set $F(\varepsilon_t | A_t^C)$ equal to a normal distribution with mean zero and variance $E(\varepsilon_t^2 | \varepsilon_t \leq 0)$. Our procedure thus consists of classifying an observation $Y_t$ as containing a flare if a positive $\varepsilon_t$ is above the threshold $q_{(1-\alpha)}^{0}$. This happens with probability $\alpha$ under the assumption that the observation contains no flare.}

We emphasize that {using Equation~\ref{threshold} as threshold at $t$, which is equivalent to the criterion}
\begin{equation}\label{ineq}
\varepsilon_t > q_{(1-\alpha)}^{0}
\end{equation}
provides a rigorous definition for identifying events in stellar light curves, which we then interpret as flares.  Using the inequality in Equation~\ref{ineq} is tantamount to specifying the probability of classifying a random fluctuation in the light curve as a flare as $\alpha$.  Thus, the proportion of events flagged as candidate flares according to this method provides an estimate of $Pr(\varepsilon_t {>} 0 \cap A_t)$, i.e.\ the probability that a flare occurs.  In practice, if $\alpha$ is set too high, the estimate will tend to be too high, and conversely, if $\alpha$ is set too low, the estimate will tend to be too low.  The typical value {on $\alpha$} used in {the} statistical literature is $\alpha=0.05$, which is the equivalent {to a $1.96\sigma$ deviation from zero in a standard normal distribution}.  In Section~\ref{sec:HB_BH} below, we discuss how to adjust $\alpha$ for the case where multiple comparisons are made.

In practice, our method proceeds as follows. First, we obtain the scaled residuals $\widehat{\varepsilon}_t := \bigr(Y_t - \widehat{\mu}(t) - {\widehat{\mu}_X(t)}\bigr) / \widehat{\sigma}_t$, where the $\widehat{\mu}(t)$'s are the fitted values of the harmonic trend in Equation~\ref{HF}, the ${\widehat{\mu}_X(t)}$'s are the fitted values of the ARMA specification in Equation~\ref{eq:arma:model}, and the $\widehat{\sigma}_t^2$'s are the fitted values of the GARCH specification in Equation~\ref{eq:garch:model} so that $\widehat{\sigma}_t = \sqrt{\widehat{\sigma}_t^2}$. 

Second, we obtain the estimate $\widehat{E}(\varepsilon_t^2 | \varepsilon_t \leq 0 \cap A_t^C)$ as $(1/n^-) \sum (\widehat{\varepsilon}_t^-)^2$. The sum in this average is taken over negative values of $\widehat{\varepsilon}_t$ only, the $\widehat{\varepsilon}_t^-$'s, and $n^-$ is the number of negative values $\widehat{\varepsilon}_t^-$. 

Third, we obtain the quantile estimate 
$$\widehat{q}_{(1-\alpha)}^{0} = \widehat{\Phi}^{-1}(1-\alpha)\,,$$ where 
$$\widehat{\Phi}(1-\alpha) = N\bigr(0, \widehat{E}(\varepsilon_t^2 | \varepsilon_t \leq 0 \cap A_t^C) \bigr)\,.$$ 

Finally, we classify the $Y_t$'s that exceed
\begin{equation}\label{eq:detect_candidate}
    \widehat{y}_{(1-\alpha)}^{0}(t) = \widehat{\mu}(t) + {\widehat{\mu}_X(t)} + \widehat{\sigma}_t \widehat{q}_{(1-\alpha)}^{0}
\end{equation}
as candidate stellar flares.

\subsubsection{Holm's Bonferroni and Benjamini-Hochberg procedure for multiple testing}\label{sec:HB_BH}

A crucial step of any detection process is to balance the detectability of true, but weak, signals against the false detections of background fluctuations.  It is not sufficient to merely check the $p$-value of the null distribution at each bin to decide on the reality of an event, because these $p$-values define the {\sl fraction} of times that a detection threshold can be exceeded as a matter of course by background fluctuations.  Thus, when a large number of such tests are carried out, such as for every bin in a light curve, corrections must be made to control for false discoveries \citep[see, e.g., discussion by][in their section 6]{2024ApJS..275...30T}.

To evaluate multiple observations simultaneously for the presence of flares, the likelihood of encountering false positives inherently increase. That is because each test carries its own probability of incorrectly rejecting the null hypothesis, and these probabilities accumulate across all tests. 
Here, we describe two methods that have been devised to account for the effect of multiple testing: Holm's Bonferroni \citep[HB;][]{Holm's} and Benjamini-Hochberg \citep[BH;][]{BenHoch} to effectively control the probability of false positives.  The BH procedure \citep[recommended by][]{2024ApJS..275...30T} targets the False Discovery Rate (FDR), which is the expected proportion of false positives among all identified flares. In other words, as shown below with simulations (Section~\ref{sec:simulation}), the BH procedure is suitable when some level of false detections is tolerable in exchange for higher sensitivity in flare detection.  The HB procedure, which targets the Family-wise Error Rate (FWER) is more conservative, and is preferred when minimizing false detections is critical, even if it means potentially missing some true flares.  Loosely speaking, HB allows us to be confident that a flare has definitely occurred at a given time, while BH describes confidence in the total number of detections.

\begin{figure*}
\centering
    \includegraphics[height=2.7in]{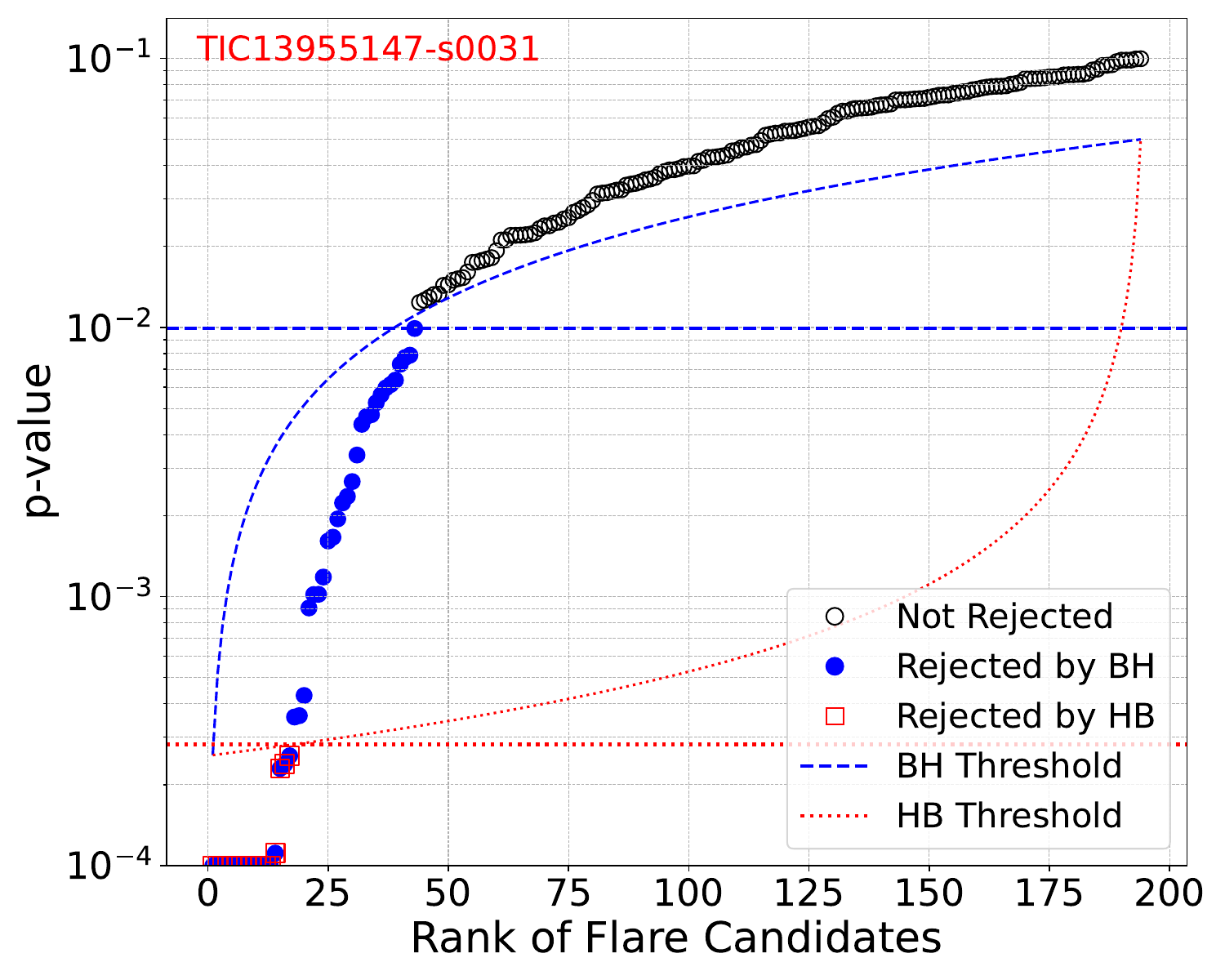}
    \includegraphics[height=2.7in]{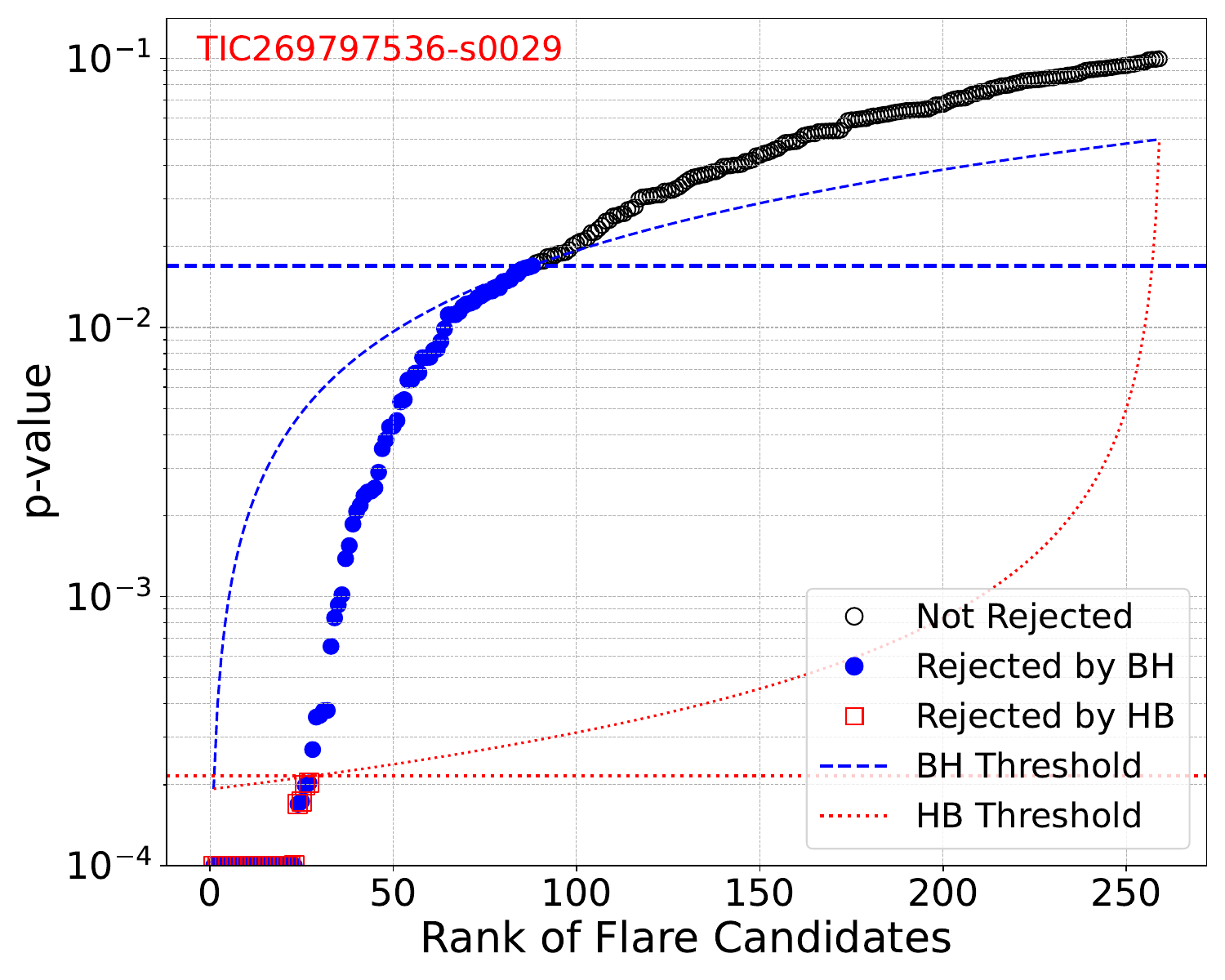}
\caption{Illustration of screening the flare detections by correcting for multiple-testing.  The two panels represent the candidate flares identified at the 95\% threshold of the distribution in $\varepsilon_t$ in Figure~\ref{fig:epsilont_dist} for \ticone-s0031 ({\sl left}) and \tictwo-s0029 ({\sl right}).  The $p$-values of each candidate flare is shown as a function of its rank order, with $p$ bounded below at {$10^{-4}$} for the sake of visibility.  The curved lines represent the estimated threshold at each rank, and the horizontal lines mark the adopted critical $p$-value thresholds.  The dashed and dotted lines represent the BH (Benjamini-Hochberg) and the HB (Holm's Bonferroni) procedures respectively.  Flare candidates that are found via BH and HB (i.e., rejected as arising from the standard distribution of $\varepsilon_t$) are marked with a filled circle and open square respectively, and candidates discarded as flares are plotted as open circles.
}
\label{fig:bh_hb_thr}
\end{figure*}

The precise execution of the BH and HB procedures are illustrated in Figure~\ref{fig:bh_hb_thr}.  First, all the possible time bins that have large enough $\varepsilon_t$ to have $p$-values smaller than a nominal detection threshold (i.e., rejected at $\alpha<0.05$, see Figure~\ref{fig:epsilont_dist} and Equation~\ref{eq:detect_candidate}) and are thus considered as possible flare candidates, are collected, and ranked in order of their $p$-values.  The corresponding BH and HB adjusted thresholds are estimated for each case, and those flare candidates that have $p$-values {\sl below} the adjusted threshold are flagged as flares.  In the illustrative cases shown in Figure~\ref{fig:bh_hb_thr}, flares found by BH while controlling for FDR are shown as blue dots, and those found by HB while controlling for FWER are shown as red squares.  In the following, we describe the process in greater detail.

As formulated in \ref{eq:detect_candidate}, each observation $Y_t$ is individually assessed for flare activity. Formally, the hypotheses for testing each time point $t$ can be articulated as
\begin{eqnarray}
\mathbf{H}_0 : & \varepsilon_t \leq q_{(1-\alpha)}^0 & \mathrm{(No~flare~occurs~at~time}~t)\\
\mathbf{H}_A : & \varepsilon_t>q_{(1-\alpha)}^0 & \mathrm{(A~flare~occurs~at~time}~t)
\end{eqnarray}

For each candidate flare, the p-values $p_t$ is calculated based on the standardized residual $\hat{\varepsilon}_t$,
\begin{equation}
p_t=1-\operatorname{erf}\left(\frac{\hat{\varepsilon}_t}{\sqrt{2} \widehat{E}(\varepsilon_t^2 | \varepsilon_t \leq 0 \cap A_t^C) }\right) \,,
\end{equation}
which is the probability of observing a residual as extreme as $\hat{\varepsilon}_t$. We sort the $p$-values $p_t$ in ascending order to obtain $p_{(1)} \leq p_{(2)} \leq \cdots \leq p_{(m)}$, with $\mathbf{H}_{(i)}$ being the associated null hypothesis of $p_{(i)}$, where $m$ is the number of the total hypothesis tests.

HB procedure is a step-down procedures that starts with the smallest p-value but applies progressively more stringent thresholds as it "steps down" the ranked list. Specifically, each $p$-value is compared to a threshold that becomes stricter for larger ranks. In other words, the procedure examines $p_{(i)}$ from $i=1$ and stops when it encounters the first $i$ (denoted by $i_{HB}$) such that 
\begin{equation}
p_{(i)}>\alpha /(m-i+1) \,.
\end{equation}
rejecting all $\mathbf{H}_{(j)}$ with $j=1,2, \ldots, i_{HB}-1$.  This ensures that the probability of making one or more false discoveries remains below a predefined significance level $\alpha$. This procedure controls the Family Wise Error Rate (FWER), and is less conservative than the traditional Bonferroni correction, as it sequentially tests hypotheses from the most to the least significant. Consequently, it allows greater sensitivity in detecting true flares while maintaining stringent error control. 

BH procedure, on the other hand, is a step-up procedure begins with the smallest p-value and sequentially compares each p-value to an increasingly lenient threshold, allowing for the rejection of hypotheses as it "steps up" through the ranked list. It finds the largest $i$ (denoted by $i_{BH}$) such that
\begin{equation}
    p_{(i)} \leq \frac{i \alpha}{m}
\end{equation}
and rejects all $\mathbf{H}_{(j)}$ with $j=1,2, \ldots, i_{BH}$.

\section{Simulation Results: Detection Power by Injection}
\label{sec:simulation}

\begin{figure}
\includegraphics[width=\columnwidth]{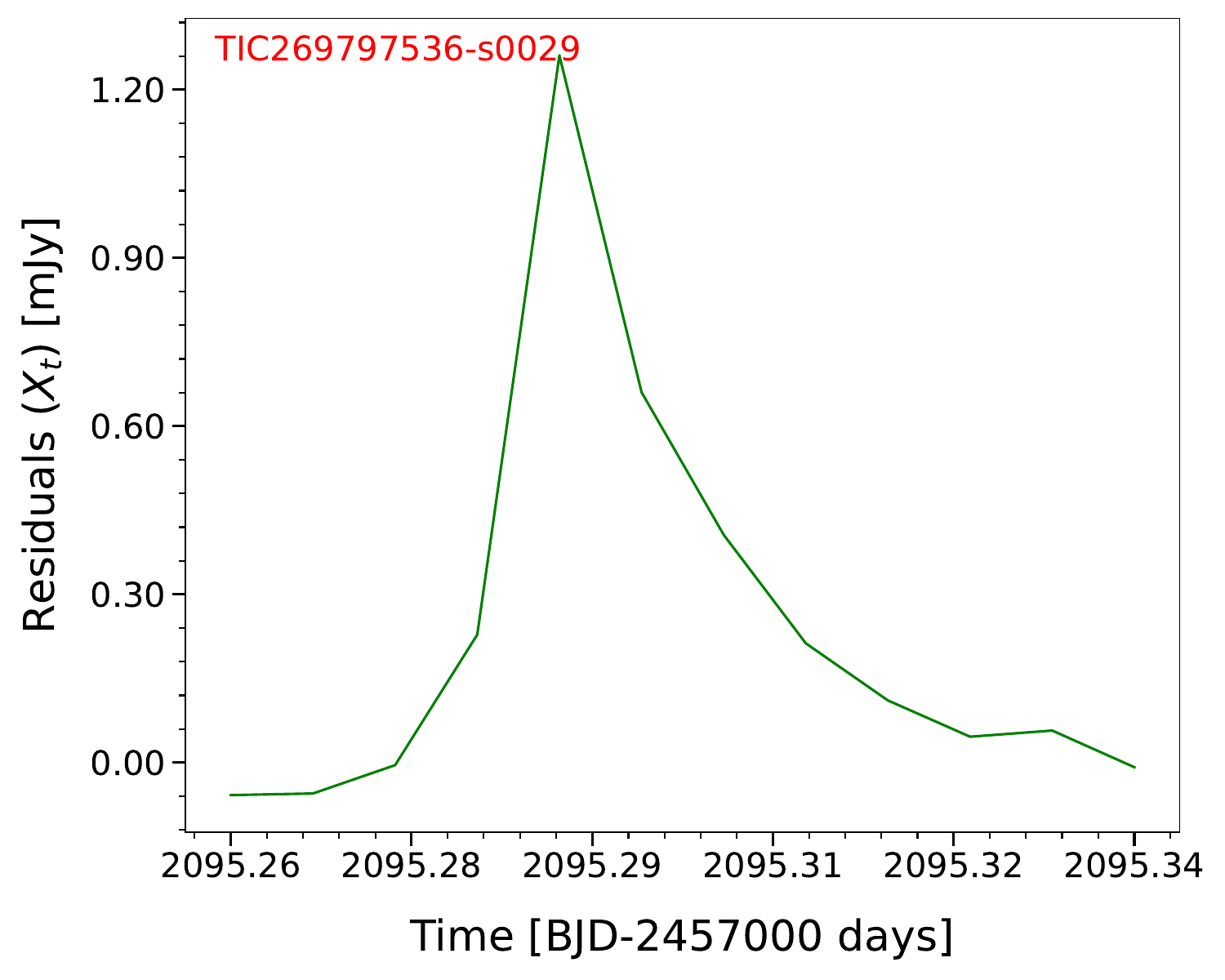}
    \caption{Flare template used for injection in simulations.  This flare appears in the first span of \tictwo~Sector-29, and is injected into harmonic-detrended and flare-free representations of light curves after rescaling the peak flux and translating (see text).
    }
    \label{fig:inject_temp}
\end{figure}

\begin{figure*}
\centering
    \includegraphics[height=2.8in]{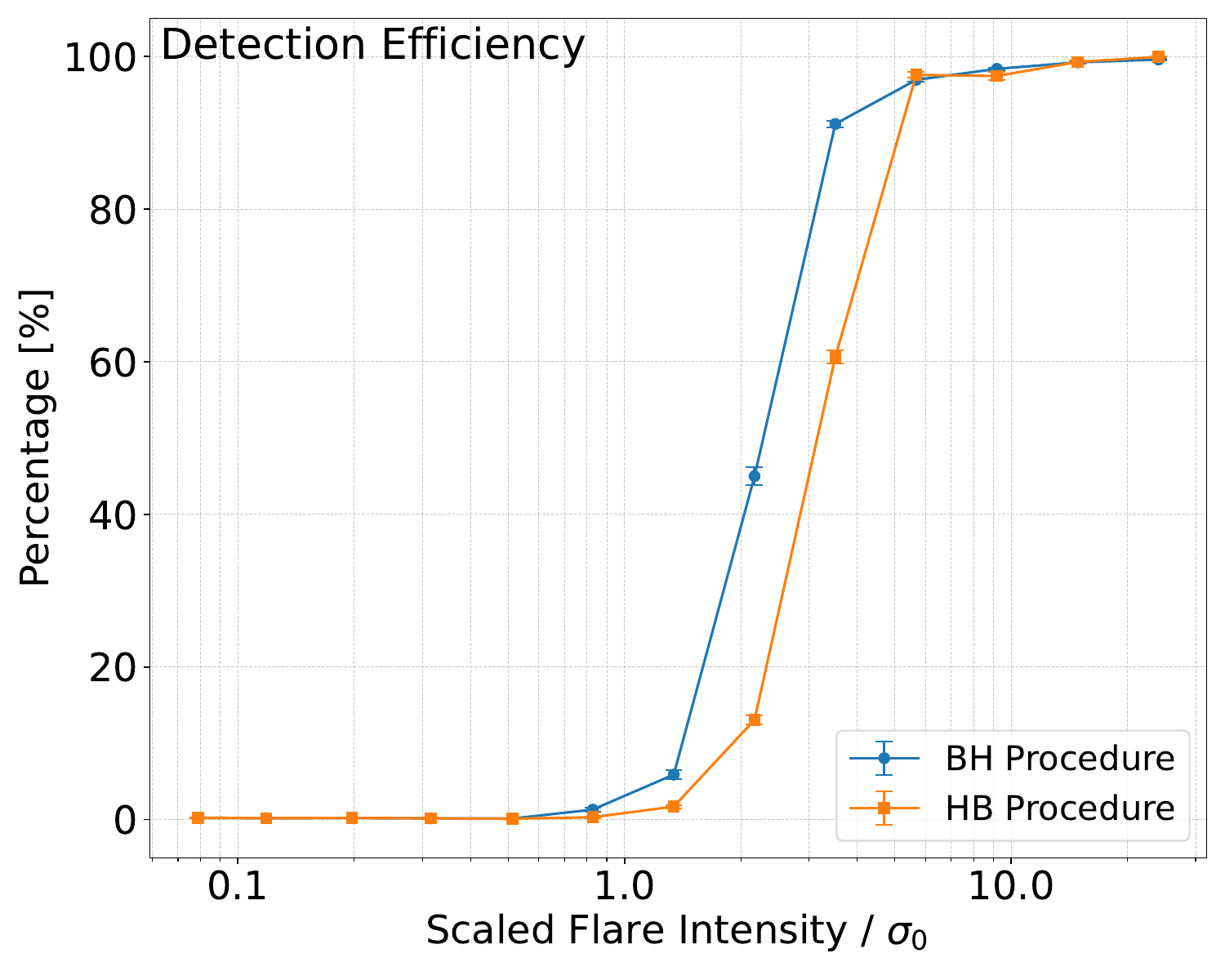}
    \includegraphics[height=2.8in]{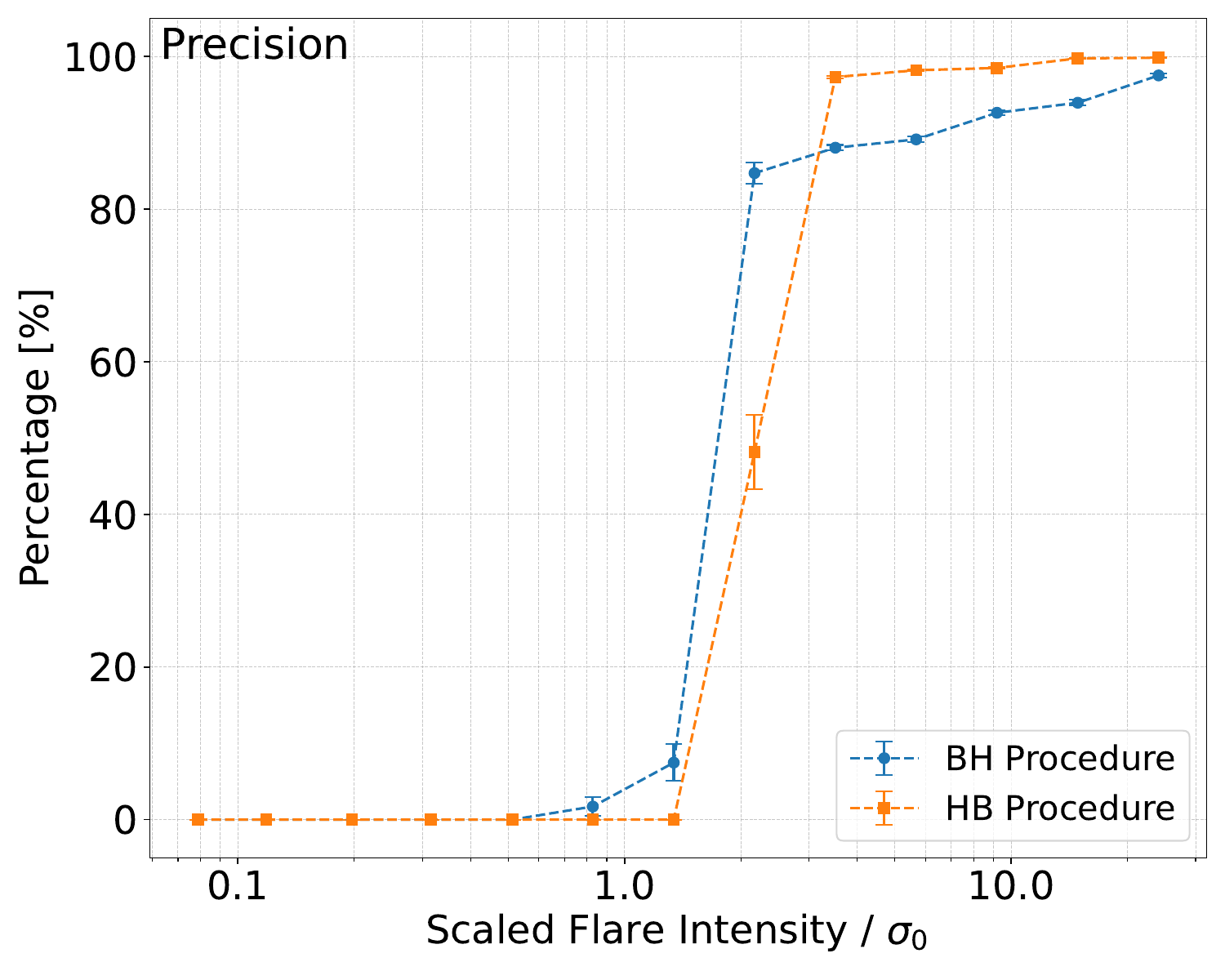}
    \caption{
    Characterizing flare detectability via injecting flares into a harmonic-detrended and signal-free light curve, illustrated for data from \tictwo~Sector~29.  The left panel shows the Detection Efficiency (the fraction of injected flares that are detected), and the right panel shows the Precision (fraction of true positive detections), as a function of a multiplicative scaling factor applied to a template (see Figure~\ref{fig:inject_temp}) relative to the local noise present in the light curve.  The two curves denote the different multiple-test corrections applied, for the HB (red filled squares) and the BH (blue filled circles).
    }
\label{fig:injection_exp}
\end{figure*}

We assess the effectiveness of our flare detection method by conducting a simulation study where we inject synthetic flares into light curves and recover them.  We select as a template a clearly discernible flare, with a peak flux of $\approx$1.2~mJy, from the first span of \tictwo~Sector~29 (see Figure~\ref{fig:inject_temp}).  The template flare is rescaled by factors ranging from $0.01\times$ to $1.0\times$ and injected into a signal-free light curve, and the detection process is run to test whether the injected flare is recovered.  We construct the signal-free light curve by first doing harmonic detrending and then excising all candidate flares.  For each rescaling factor, 100 template flares are injected into the resulting noise-only light curve at a regular spacing.  We run the modified light curves through our full analysis framework, which includes ARMA and GARCH modeling to account for autocorrelation and heteroskedasticity.  We then apply the BH and HB thresholding procedures, both at $\alpha=0.05$, to detect the injected flares. By comparing the detected flares with the known injection locations, we calculate the detection efficiency (the fraction of injected flares that are detected, Equation~\ref{eq:detectrate}), and precision (the fraction of injected flares among all detections, see Equation~\ref{eq:precision}) for each scaling factor.  Error bars are computed from repeated application of the procedure ($50\times$) with the location of the first injection varied randomly by 10\% of the observation duration.  The results are illustrated for the case of \tictwo-Sector~29 in Figure~\ref{fig:injection_exp}.  The results of the simulations for other observations are similar.

For each scaling factor $\mathcal{S}$, we denote the set of injected flares to be $\mathcal{I}_F^\mathcal{S}=\left\{\mathcal{F}_1^\mathcal{S}, \mathcal{F}_2^\mathcal{S}, \ldots, \mathcal{F}_{N_F}^\mathcal{S}\right\}$, where each $\mathcal{F}_i^\mathcal{S}$ is the index region of the $i$-th injected flare at scaling factor $\mathcal{S}$. After processing the data through the analysis framework and applying the detection methods, we obtain the set of detected flares $\mathcal{I}_D^\mathcal{S}=\left\{\mathcal{D}_j^\mathcal{S}, \ldots, \mathcal{D}_{N_D^\mathcal{S}}^\mathcal{S}\right\}$, where each $\mathcal{D}_j^\mathcal{S}$ is the index region of the $j$-th detected flare at scaling factor $\mathcal{S}$. The total number of injected flares for each scaling factor is $N_F = 100$.

We define the True Positives $\left(\mathrm{TP}_\mathcal{S}\right)$ as the number of injected flares correctly detected at scaling factor $\mathcal{S}$, while the False Positives $\left(\mathrm{FP}_\mathcal{S}\right)$ as the number of detected flares that do not correspond to any injected flare at scaling factor $\mathcal{S}$,
\begin{eqnarray}
\mathrm{TP}_\mathcal{S}&=&\left|\left\{\mathcal{F}_i^\mathcal{S} \in \mathcal{I}_F^\mathcal{S} \mid \exists \mathcal{D}_j^\mathcal{S} \in \mathcal{I}_D^\mathcal{S} \mathrm{such~that}~\mathcal{F}_i^\mathcal{S} \cap \mathcal{D}_j^\mathcal{S} \neq \emptyset\right\}\right| \nonumber\\
\mathrm{FP}_\mathcal{S}&=&\left|\left\{\mathcal{D}_j^\mathcal{S} \in \mathcal{I}_D^\mathcal{S} \mid \forall \mathcal{F}_i^\mathcal{S} \in \mathcal{I}_F^\mathcal{S}, \mathcal{D}_j^\mathcal{S} \cap \mathcal{F}_i^\mathcal{S} = \emptyset\right\}\right|
\label{eqn:TPFP_sim}
\end{eqnarray}

We focus on two off-the-shelf metrics, namely {\sl Detection Efficiency} and {\sl Precision}. The Detection Efficiency measures the proportion of injected flares correctly detected at scaling factor $\mathcal{S}$,
\begin{equation}
    \operatorname{DetectionEfficiency}(\mathcal{S}) = \frac{\mathrm{TP}_\mathcal{S}}{N_F}
    \label{eq:detectrate}
\end{equation}

However, Detection Efficiency alone measures only the proportion of actual flares correctly identified (also known as the sensitivity) without accounting for false positives.  Precision complements this by indicating the proportion of detected flares that are true positives.  We denote the Precision at scaling factor $\mathcal{S}$ as:
\begin{equation}
    \operatorname{Precision}(\mathcal{S}) = \frac{\mathrm{TP}_\mathcal{S}}{\mathrm{TP}_\mathcal{S} + \mathrm{FP}_\mathcal{S}}
    \label{eq:precision}
\end{equation}

The results of the flare injection study are shown in Figure \ref{fig:injection_exp}, as a function of scaling factors, themselves scaled by the noise in the light curves.  The scatter present in the light curves after detrending by the harmonic-component and removing candidate flares represents the noise $\sigma_0$, which controls detectability.  Notice that the Detection Efficiency drops to 50\% between 2 and 3 $\mathcal{S}/\sigma_0$.  For \tictwo-Sector~31, $\sigma_0\approx{0.04}$~mJy (see Table~\ref{tab:flarecensus} below), which corresponds to a peak flux of 
$\approx$0.1~mJy % 2*1.22*0.041
$\approx$0.15~mJy % 3*1.22*0.041
for BH and HB respectively.  Weaker flares become difficult to detect below these values.  Additionally, the Precision drops below 50\% at $\approx$2~$\mathcal{S}/\sigma_0$, at $\approx$0.1~mJy, suggesting that a flare detected at peak flux at this level is as likely to be a false positive as a true positive.  Fundamentally, the detection sensitivity of stellar flares in TESS light curves decreases rapidly, and is unreliable, for peak fluxes $<1.22{\cdot}2\sigma_0$~mJy.  For $\sigma_0{\approx}0.04$~mJy, this corresponds to a limiting sensitivity of ${\approx}2{\times}10^{-13}$~erg~s$^{-1}$~cm$^{-2}$.

We further observe that the detection efficiency for the BH procedure increases more rapidly and reaches 100\% at lower scaling factors compared to HB method. This means that as the flare intensity $(\mathcal{S})$ increases, the BH procedure is more sensitive and detects flares more readily. On the other hand, although Holm's method is more conservative in terms of detection efficiency, it rarely produces false positives, resulting in precision values of either 0 or 100\%. Conversely, the BH procedure eventually includes some false positives as the intensity of the injected flares increases, stabilizing around a precision level of $95\%$ consistent with its control of the false discovery rate at the 5\% level \citep{BenHoch}. The results align with our expectation that the BH procedure allows for a controlled proportion of false positives to achieve higher detection efficiencies, while HB method strictly controls the family-wise error rate, leading to fewer detections but higher precision \citep{Holm's}.

\section{Application}\label{sec:apply}

\begin{figure}
    \centering
    \includegraphics[width=0.45\linewidth]{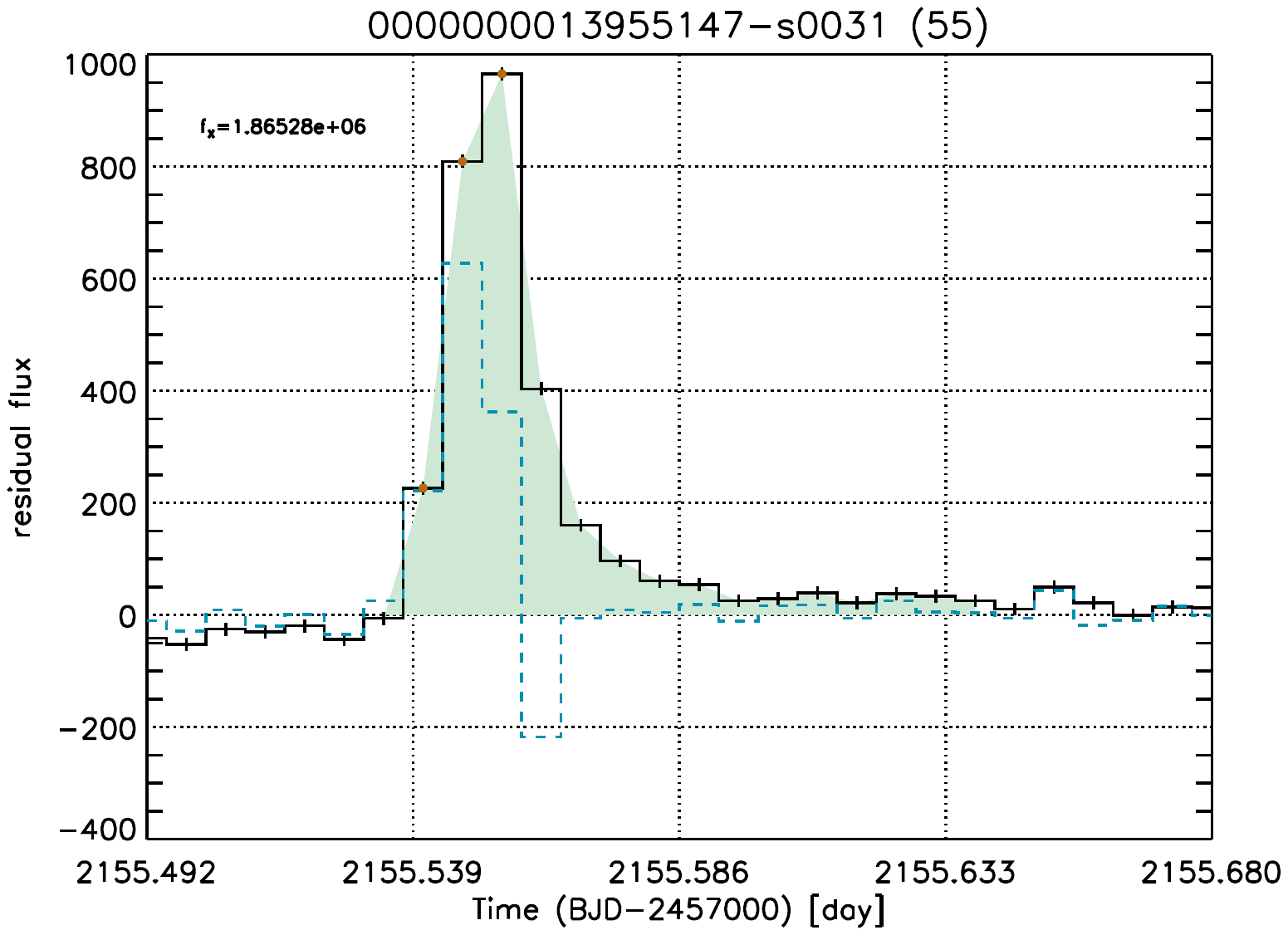}
    \includegraphics[width=0.45\linewidth]{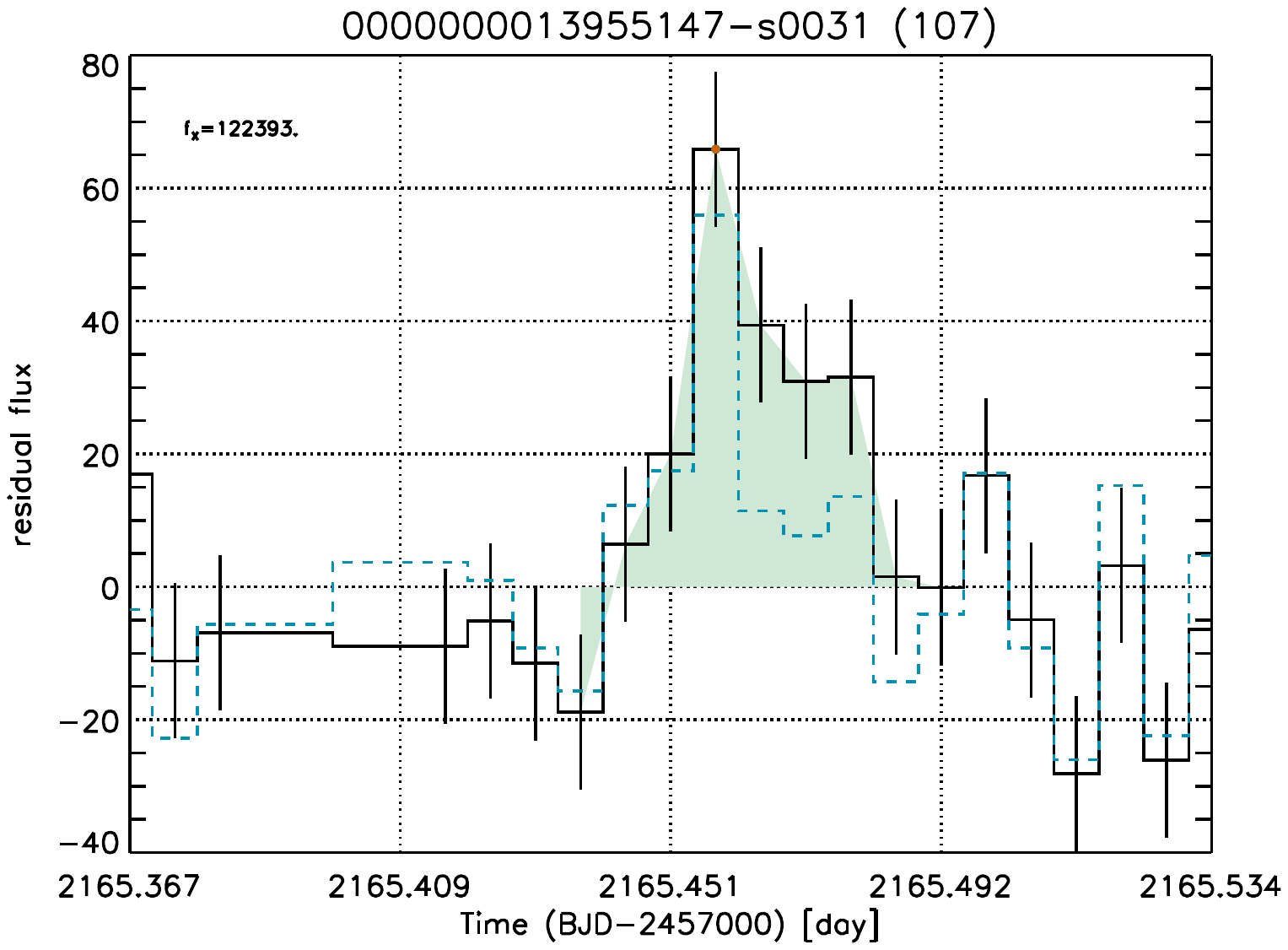}
    \includegraphics[width=0.45\linewidth]{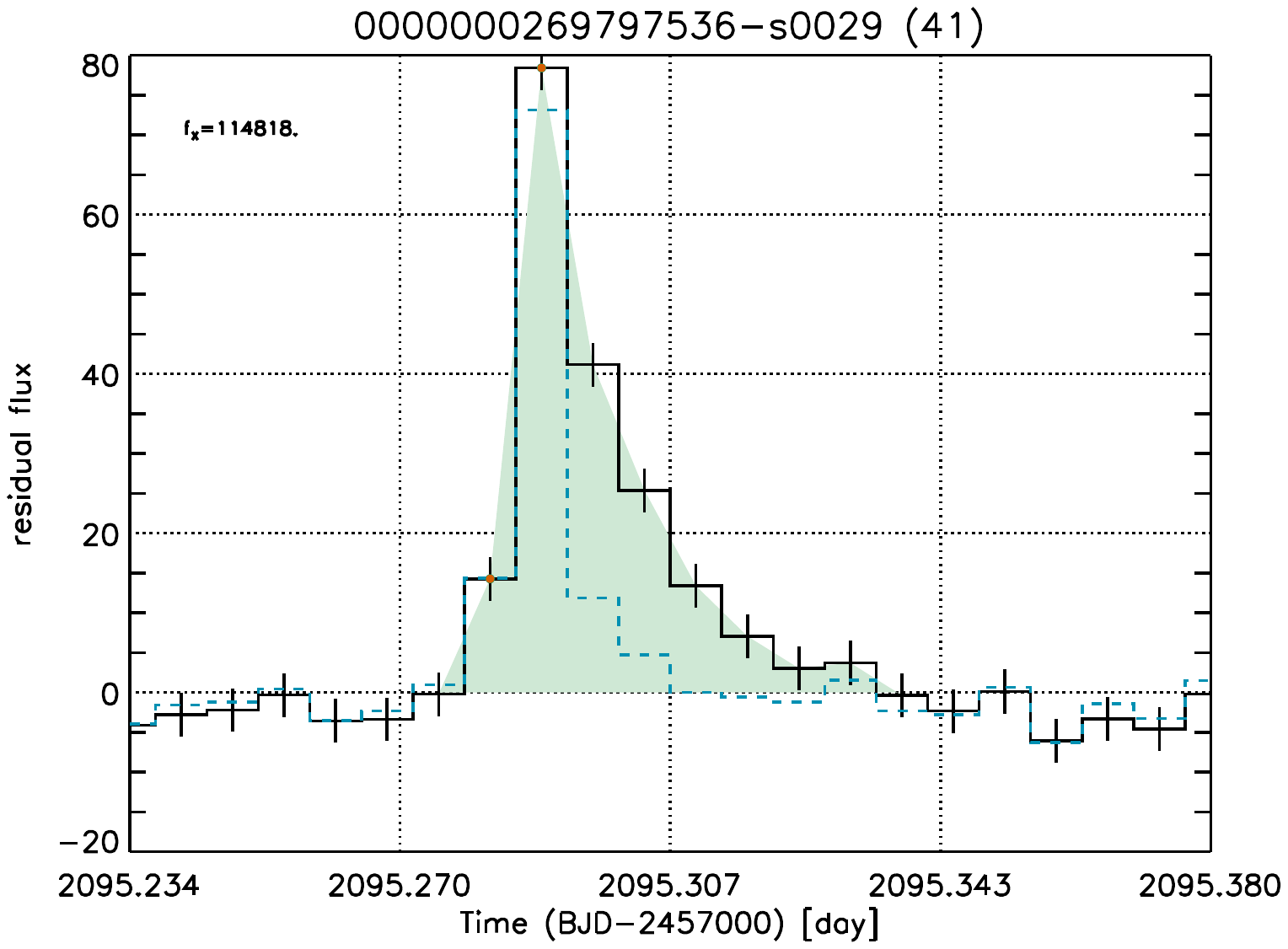}
    \includegraphics[width=0.45\linewidth]{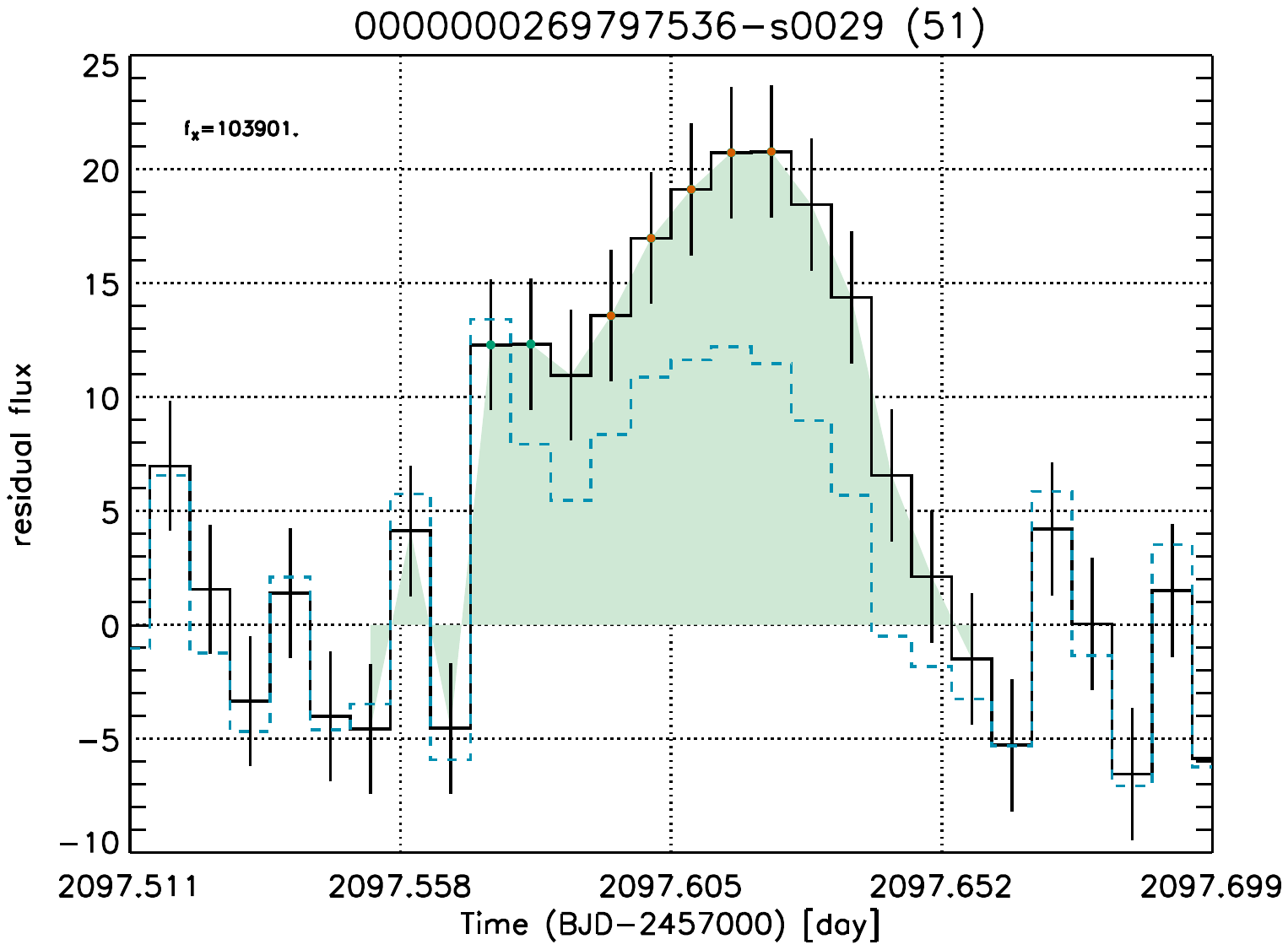}
    \caption{{Illustrative examples of detected flares and flux measurements.  Each panel shows a short time interval zoomed into a detected flare for \ticone\ (top) and \tictwo (bottom), for a relatively large and simple (left) and small and complex (right) flare.  The haromic detrended light curve is shown as the black histogram with $1\sigma$ error bars, and the shaded region represents the area under the light curve that is used to estimate the flare energy $E$.  The dashed line is the ARMA-GARCH residuals $Z_t$ that has accounted for the correlations present in the data, and thus marks the occurrence of flares as large deviations (as leveraged for detection in Figure~\ref{fig:epsilont_dist}). Note that the flare shown at lower left is the same one as was used as the template for injection in the simulations (Figure~\ref{fig:inject_temp}).}}
    \label{fig:fittedflares}
\end{figure}

We apply our method to all contiguous pieces of the sector light curves listed in Table~\ref{tab:stars}.  After determining which time bins in the detrended light curves contain flare candidates, we first merge those candidates which are within three time bins of each other into a single flare.  We then extend the flare interval in both directions until the detrended intensities drop below zero.  However, if another candidate flare occurs before the intensity drops below zero, we stop the flare interval at the deepest minimum between the two flare candidates (none of these happen to be flares eventually selected under either the HB or BH schemes).  We then compute the energy present in each time bin as the product of the residual flux (after harmonic detrending) and the time bin width, and sum up the energies within the interval to compute the total energy $E$ released in the event.  We also record the maximum value within the flare interval of the detrended residuals as the peak flux $P$ of the flare.  {We show representative examples of how this process works in Figure~\ref{fig:fittedflares}.  The flare energies are computed over the shaded intervals.  The figures also show the fitted ARMA-GARCH residuals $Z_t=\sigma_t\varepsilon_t$ (dashed histogram).  Notice that flares are detected when there are large deviations in $Z_t$, as times when the data are incompatible with the autocorrelation structure in the light curve.  The ARMA-GARCH modeling thus isolates instances when there are impulses that stand out beyond the usual fluctuations, and can, in principle, probe flare onset.  We flag the peak in $\varepsilon$ where a flare is detected as the impulsive energy $I$, and analyze its distribution similarly to $E$ and $P$.}

If the flare candidate is flagged as a flare above either the HB or BH threshold, it is kept for further analysis; otherwise it is discarded.  A census of the detected flares is in Table~\ref{tab:flarecensus}.  As expected, more flares are identified under the BH threshold than the HB threshold: 145 and 58 respectively for \ticone, and 460 and 232 respectively for \tictwo.  Note that the residual noise in the detrended light curves, $\sigma_0$ is systematically lower by a factor $1.7\times$ for the 30~minute cadence data than for the 10~minute cadence data, because the longer integration times serve to reduce noise by a factor of $\sqrt{3}$.  The brightness of the faintest detected flares are consistent with $\sigma_0$ (see assessments of Detection Efficiency and Precision in Section~\ref{sec:simulation} and Figure~\ref{fig:injection_exp}).
However, there are considerable variations in the numbers and characters of the detected flares over time for the same star.  The peak luminosities in the different Sectors range between $6{\times}10^{29} - 2{\times}10^{31}$~erg~s$^{-1}$ for \ticone, and $7{\times}10^{28} - 1.2{\times}10^{31}$~erg~s$^{-1}$ for \tictwo.  The flare rate varies between $0.4-1$~day$^{-1}$ (for HB threshold) and $0.7-2.6$~day$^{-1}$ (for BH threshold) for both stars; the variations between Sectors is larger than the difference between the stars.  Additionally, in \ticone, Sector~32 has $\approx\frac{2}{3}\times$ more flares than Sector~31, the latter which has a flare that is $\approx4\times$ brighter; and similarly in \tictwo, Sector~39 has the largest number of flares, but also predominantly weak flares, while the immediately preceding Sector~38 has the brightest flare.  This suggests that on both stars flaring is dominated by a few active regions which evolve significantly from one epoch to another.  We discuss the implications of the distributions of flare energies and peak brightness in greater detail below in Section~\ref{sec:flaredistrib}.

\begin{table*}
    \centering
    \caption{{Census summary of detected flares in each sector.}}
    \begin{tabular}{c|c|cc|c|cc|cc|c}
    \hline\hline
    sector & $\sigma_0^\dag$ & \multicolumn{2}{c}{Detected Flares} & Duration & \multicolumn{2}{c}{Flare rate [day$^{-1}$]} & \multicolumn{2}{c}{Faintest$^\ddag$} & Brightest$^\ddag$ \\
    \hfil & [mJy] & (HB) & (BH) & [day] & (HB) & (BH) & (HB) & (BH) & \hfil \\
    \hline
    \multicolumn{10}{l}{\ticone} \\
    \hline
    4 & 0.144 & 8 & 13 & 19.42 & 0.41 & 0.67 & 0.73 & 0.25 & 3.2 \\
    5 & 0.133 & 13 & 30 & 23.92 & 0.54 & 1.25 & 0.40 & 0.28 & 2.2 \\
    31 & 0.243 & 16 & 37 & 23.28 & 0.69 & 1.59 & 0.54 & 0.48 & 15.5 \\
    32 & 0.223 & 28 & 65 & 24.51 & 1.14 & 2.65 & 0.62 & 0.54 & 4.3 \\
    (all) & \hfil & 65 & 145 & 91.13 & 0.71 & 1.60 & 0.40 & 0.25 & 15.5 \\
    \hline
    \multicolumn{10}{l}{\tictwo} \\
    \hline 
    2 & 0.028 & 19 & 28 & 25.54 & 0.74 & 1.10 & 0.08 & 0.06 & 1.2 \\
    5 & 0.033 & 12 & 20 & 23.85 & 0.50 & 0.84 & 0.11 & 0.06 & 7.8 \\
    8 & 0.026 & 16 & 17 & 16.85 & 0.95 & 1.01 & 0.06 & 0.06 & 2.0 \\
    11 & 0.026 & 18 & 25 & 19.88 & 0.91 & 1.26 & 0.09 & 0.06 & 1.7 \\
    12 & 0.029 & 12 & 21 & 21.42 & 0.56 & 0.98 & 0.09 & 0.07 & 1.7 \\
    13 & 0.032 & 23 & 34 & 25.35 & 0.91 & 1.34 & 0.06 & 0.06 & 4.5 \\
    27 & 0.052 & 18 & 50 & 22.25 & 0.81 & 2.25 & 0.15 & 0.06 & 3.7 \\
    28 & 0.043 & 22 & 41 & 19.55 & 1.13 & 2.10 & 0.12 & 0.10 & 4.5 \\
    29 & 0.041 & 21 & 48 & 20.98 & 1.00 & 2.29 & 0.13 & 0.10 & 1.3 \\
    32 & 0.046 & 28 & 44 & 24.30 & 1.15 & 1.81 & 0.17 & 0.11 & 7.2 \\
    35 & 0.046 & 14 & 28 & 19.20 & 0.73 & 1.46 & 0.16 & 0.09 & 3.8 \\
    38 & 0.045 & 20 & 37 & 25.28 & 0.79 & 1.46 & 0.15 & 0.09 & 10.6 \\
    39 & 0.045 & 29 & 67 & 26.92 & 1.08 & 2.49 & 0.14 & 0.10 & 0.9 \\
    (all) & \hfil & 252 & 460 & 291.36 & 0.71 & 1.58 & 0.06 & 0.06 & 10.6 \\
    \hline
    \multicolumn{10}{l}{{\ticfour~(\aumic)}} \\
    \hline
    1 & 1.782 & 78 & 203 & 25.97 & 3.00 & 7.82 & 2.85 & 0.32 & 151.4 \\
    27 & 1.658 & 71 & 199 & 23.34 & 3.04 & 8.53 & 3.52 & 0.55 & 163.6 \\
    (all) & \hfil & 149 & 402 & 49.30 & 3.02 & 8.17 & 2.85 & 0.32 & 163.6 \\
    \hline
    \multicolumn{10}{l}{$\dag$: Noise estimated after harmonic detrending and excluding all candidate flares.} \\
    \multicolumn{10}{l}{$\ddag$: The minimum and maximum peak flux among the detected flares, in [mJy].} \\
    \end{tabular}
    \label{tab:flarecensus}
\end{table*}

\section{Discussion}\label{sec:discuss}

\subsection{Model Selection}\label{ARFIMA}
Before fitting our model Equation~\ref{eqn:arma} to the selected lightcurves, we considered the general class of ARFIMA models. The Autoregressive Fractionally Integrated Moving Average (ARFIMA) model extends the classical ARIMA model \citep[which has been applied in planet search studies, e.g.,][]{2019AJ....158...57C,2019AJ....158...58C} by incorporating a fractional differencing parameter, enabling the model to capture both short-term autocorrelation (ARMA) and long-range dependence (Fractionally Integrated). This capability makes the ARFIMA model particularly valuable for analyzing data sets where phenomena whose memory (i.e., the serial correlation)  persist over time. The ARFIMA($r,d,s)$ model is expressed as
\begin{equation}\label{eq:arfima}
\phi(B)(1-B)^d X_t=\theta(B) Z_t,
\end{equation}
where the filters $\phi(B)$ and $\theta(B)$ have been defined in Section~\ref{NPD}, and $d\in (0,\tfrac 12)$ is the order of fractional differencing.
The filter $(1-B)^d$ removes both trend (by means of the ARIMA filter $1-B$) and long-memory (via the fractional parameter $d$) from the observed light curve $X_t$. 

When $d$ is a positive integer, %we say that 
$X_t$ follows an ARIMA model, which is suitable for integrated (or non-stationary) data sets. For our light curves we have $d=0$, because the roots of the polynomials lie outside the unit circle. 
This means that we don't need to fit an ARIMA or an ARFIMA model to our data because our light curves are stationary and do not exhibit long memory. Although our data are stationary, they are conditionally heteroskedastic (viz.\ Section~\ref{sec:intro} above).

ARFIMA and GARCH modeling has been previously considered by \cite{Stanislavsky2019} in an astronomical setting, applied to disk-integrated solar soft X-ray light curves.  They show that in the periods of high-solar activity, the energy distribution of soft X-ray solar flares is well described by an ARFIMA-GARCH model. Since the X-ray emission clearly exhibits a long-range dependence often possibly heavy-tailed, the ARFIMA part of the model is used to remove trend and long-memory. The GARCH part of the model explains heteroscedastic effects of the X-ray time series process typically observed in the form of clustering of volatilities.
In addition to our method being applied to optical band stellar light curves, our approach differs from theirs on three counts:
First, we account for the time-varying trends in both amplitude and period with our harmonic detrending analysis;
Second, we find it is not necessary to carry out an ARFIMA fit to our data as it does not exhibit long memory after removing the harmonic trend, thus rendering models with fractional parameters unnecessary; and
Third, we follow on from carrying out a GARCH fit to the residuals light curve to then {\sl use} the resulting heteroskedastic error $Z_t$ as a means to detect flares.

\subsection{Flare Distributions}\label{sec:flaredistrib}

It has been well established in several studies that the distributions of both total energy and peak flux are well-described by power-laws \citep[e.g.,][]{2004A&A...416..713G,Feinstein2022}, with indices $\alpha_{Z}$ (where $Z$ represents {energy $E$, peak flux $P$, or impulse $I$}),
\begin{equation}
    \frac{dN(Z)}{dZ} = k_Z \cdot E^{-\alpha_Z} \,,
    \label{eq:dNdZ}
\end{equation}
or equivalently,
\begin{equation}
    N(>Z) = k'_Z E^{-\alpha_Z+1} \,,
    \label{eq:powlaw}
\end{equation}
where $k_{Z}$ and $k'_{Z}$ are normalization parameters that determine the total number of detected flares.  Indeed, more complex distributions, with double power-laws, have also been considered to explain TESS stellar flare distributions \citep[][]{2025arXiv250515451G}.  The allure of the scale-free power-law distribution is that it is thought to be a diagnostic of the flare onset occurs, as a manifestation of a self-organized critical process \citep[][]{1991ApJ...380L..89L,2011SoPh..274..119A}.  On the Sun, typically $\alpha_E{\approx}1.8$ \citep[see review by][]{2011SoPh..274...99A} and $\alpha_P{\approx}2$ \citep[][]{2012ApJ...754..112A}.  {Our analysis also allows for the first time access to the impulsive excess energy input into the corona that marks the beginning of flares.}

\begin{figure*}
    \centering
    \includegraphics[width=0.48\linewidth]{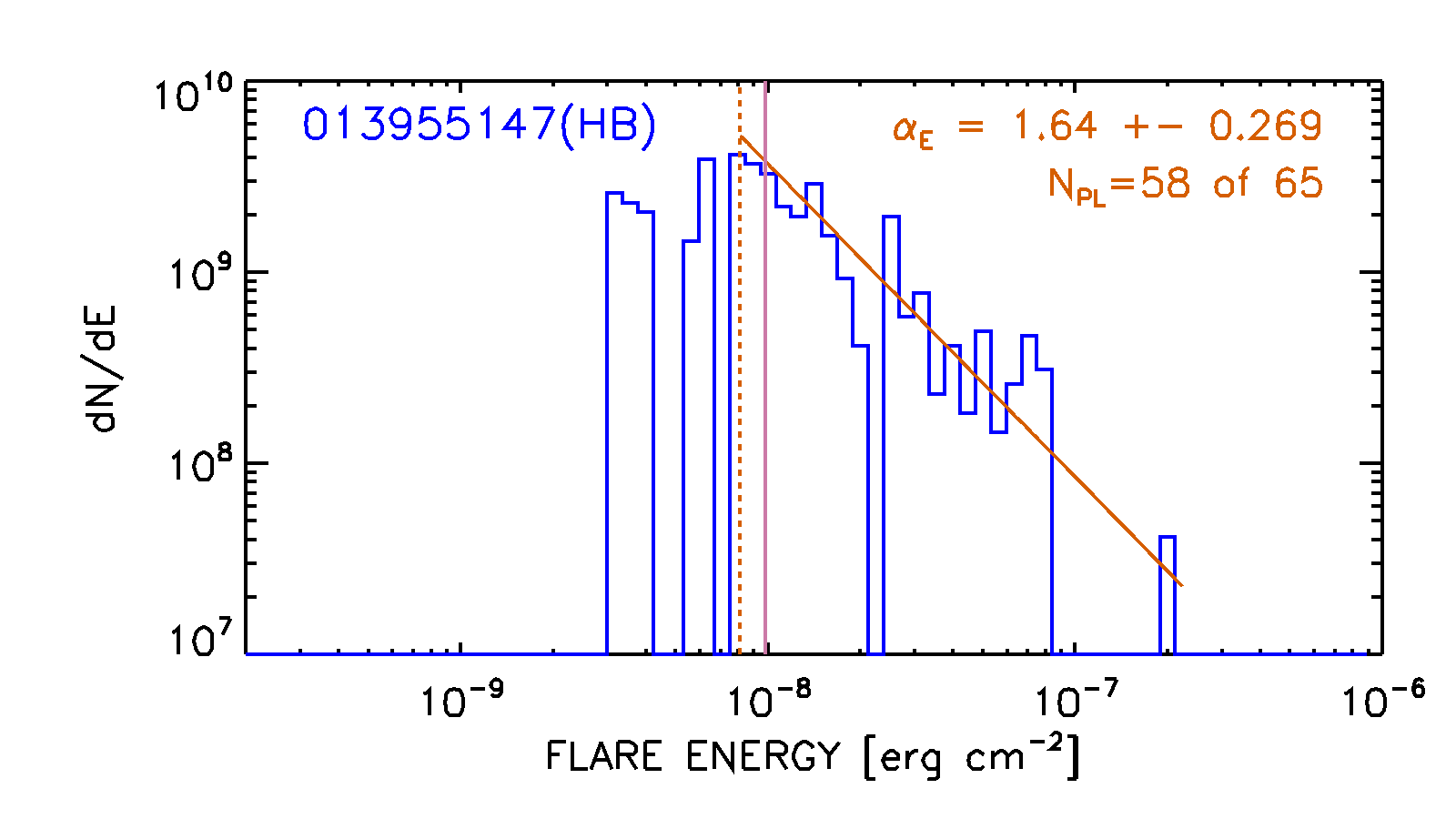}
    \includegraphics[width=0.48\linewidth]{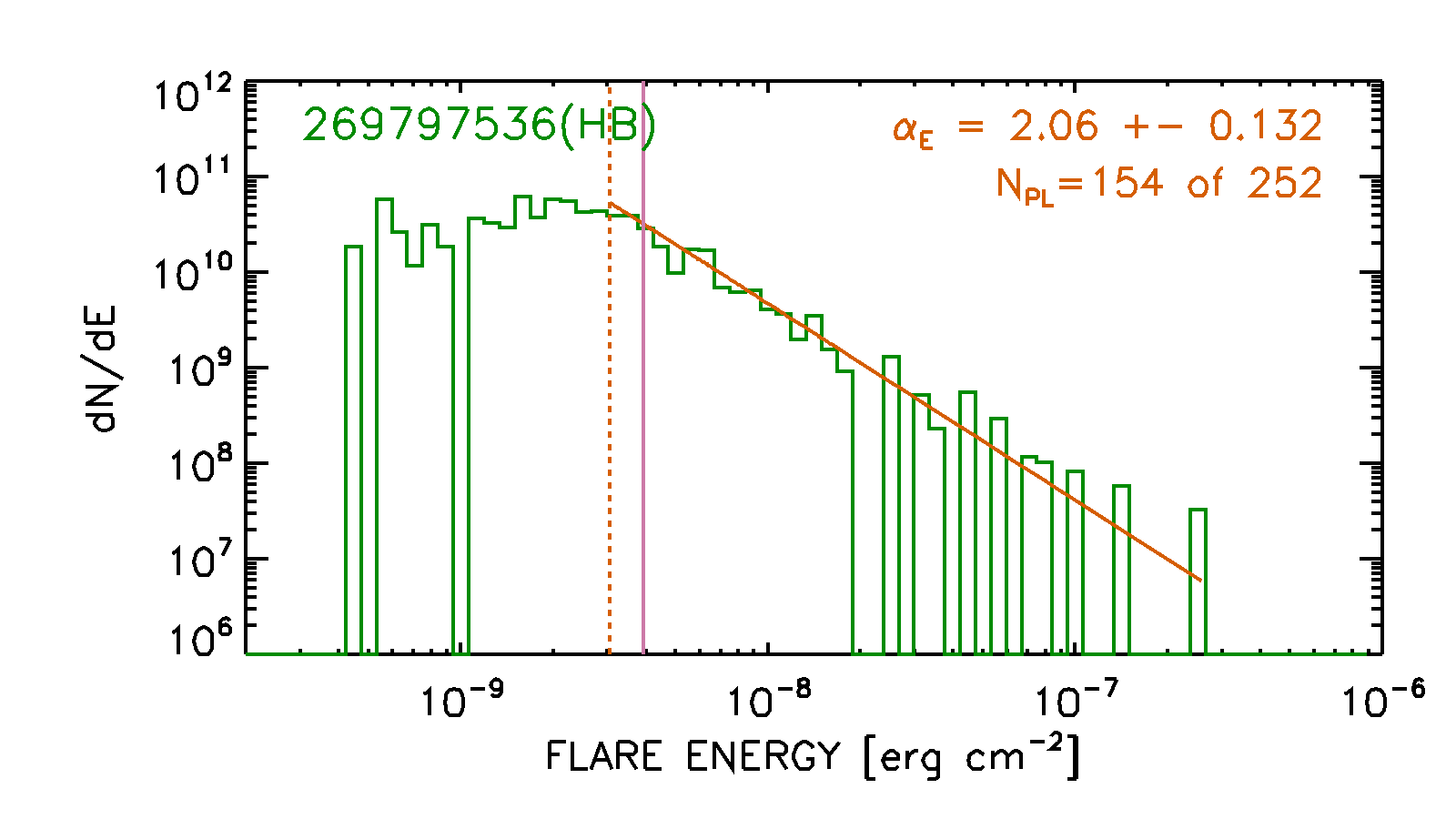}
    \includegraphics[width=0.48\linewidth]{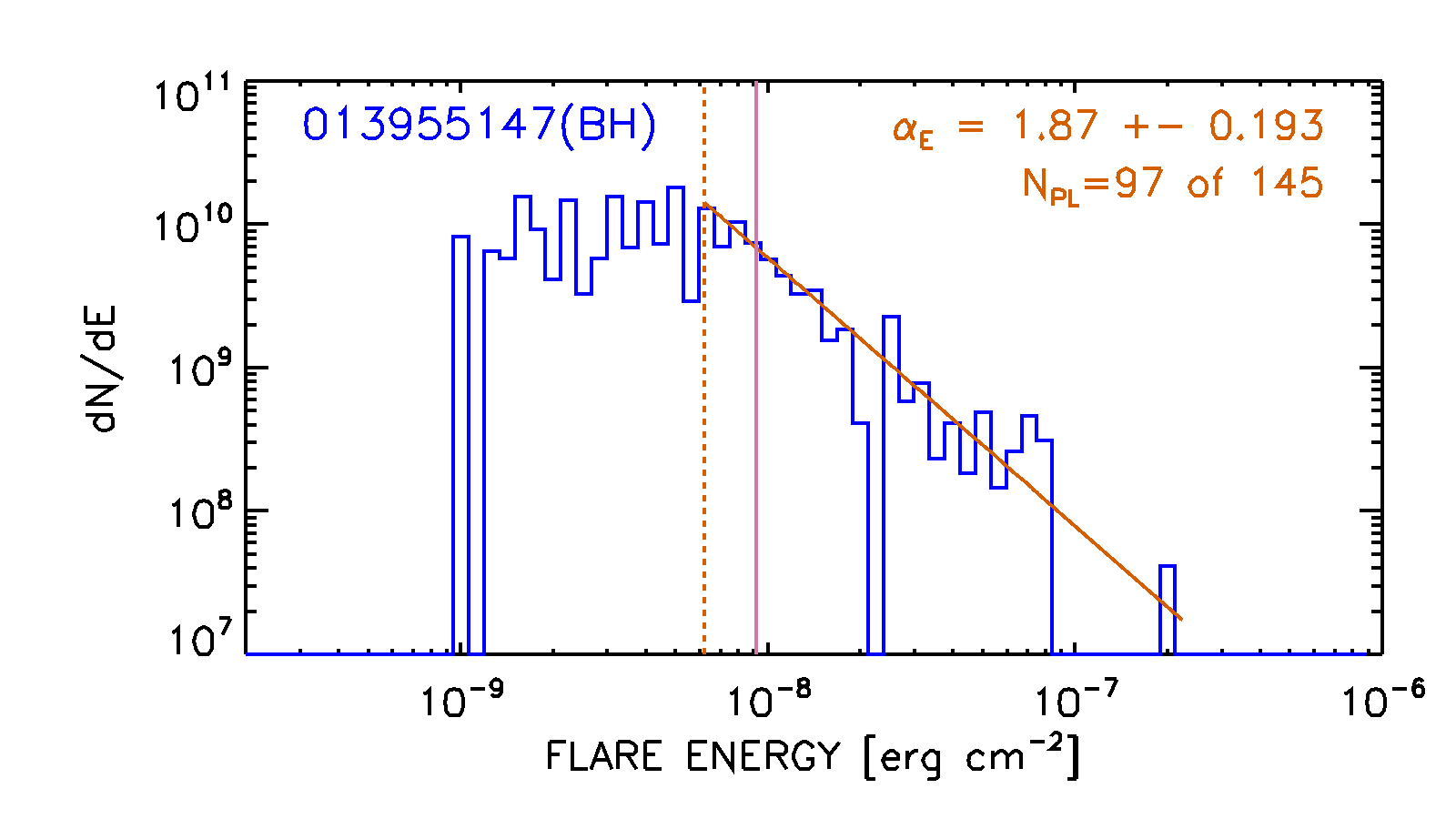}
    \includegraphics[width=0.48\linewidth]{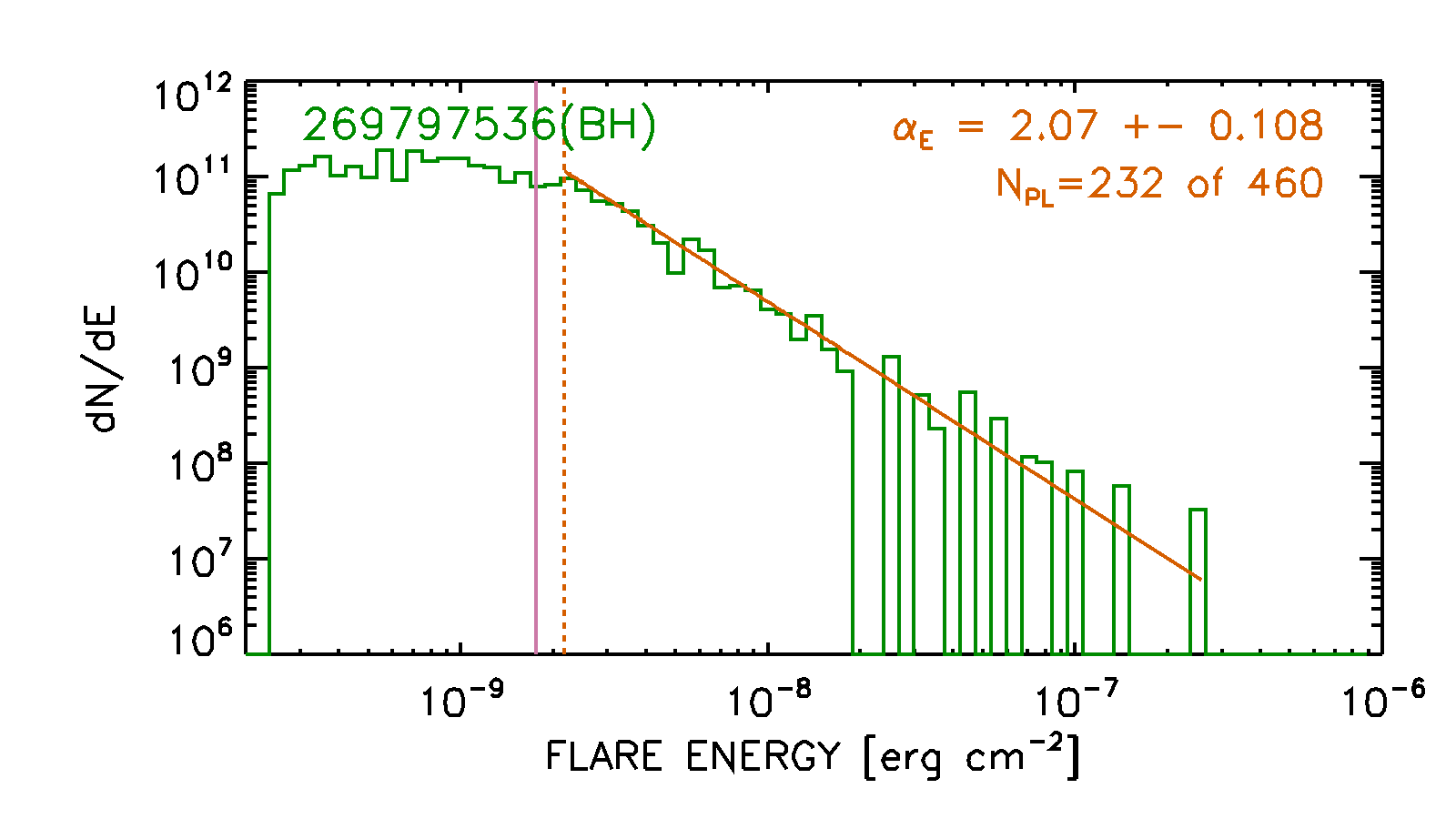}
    \caption{Flare energy distributions from the combined datasets for \ticone\ ({\sl left} column) and \tictwo\ ({\sl right} column).  The flares found using the conservative HB thresholding are shown in the {\sl top} row, and those for the less restrictive BH thresholding are shown in the {\sl bottom} row.  The observed distribution of flare energies $\frac{dN}{dE}$ is shown as the stepped histogram.  The modes of the log energy distributions are shown as the solid vertical lines, the optimal lower bounds of the power-law fits (see text) as dashed vertical lines, and the maximum likelihood power-law fits as the solid slanted lines.  The estimated power-law index $\alpha$ and the sample size used for the estimation are marked at upper right in each plot.  Notice that $\alpha$ is a solar-like $\approx{1.8}$ for \ticone, but $\gtrsim2$ for \tictwo.}
    \label{fig:dNdE}
\end{figure*}

\begin{figure*}
    \centering
    \includegraphics[width=0.48\linewidth]{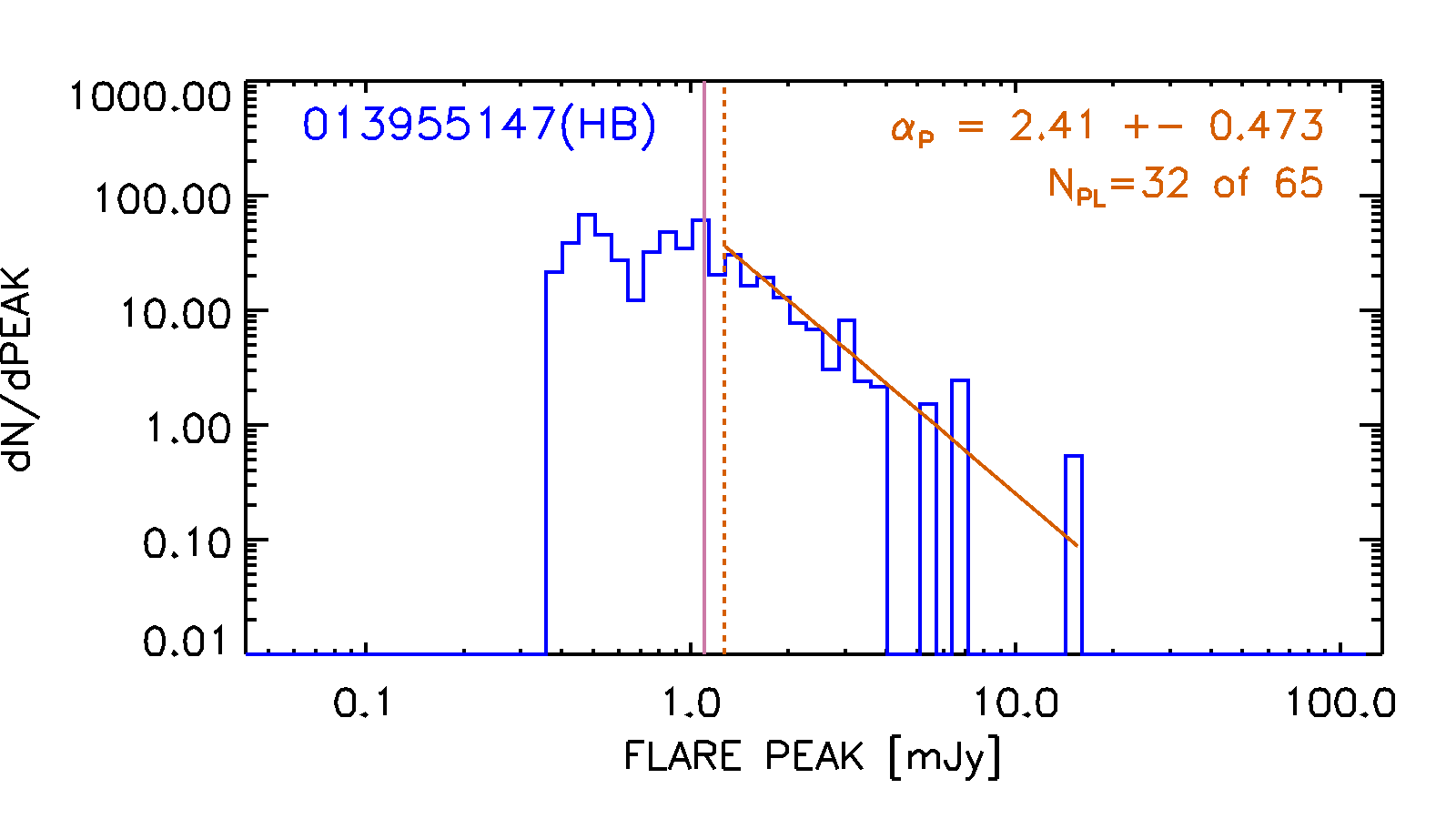}
    \includegraphics[width=0.48\linewidth]{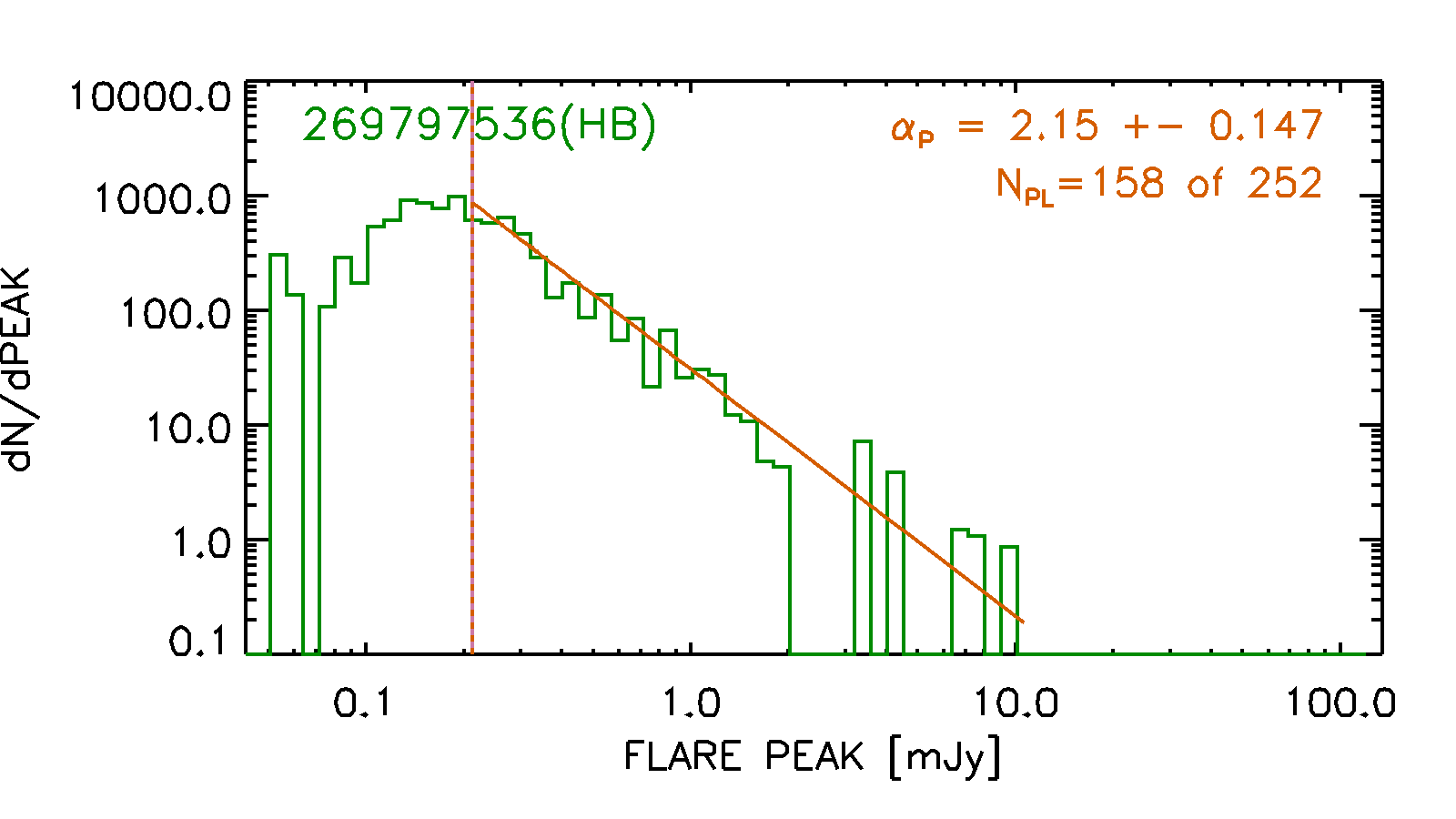}
    \includegraphics[width=0.48\linewidth]{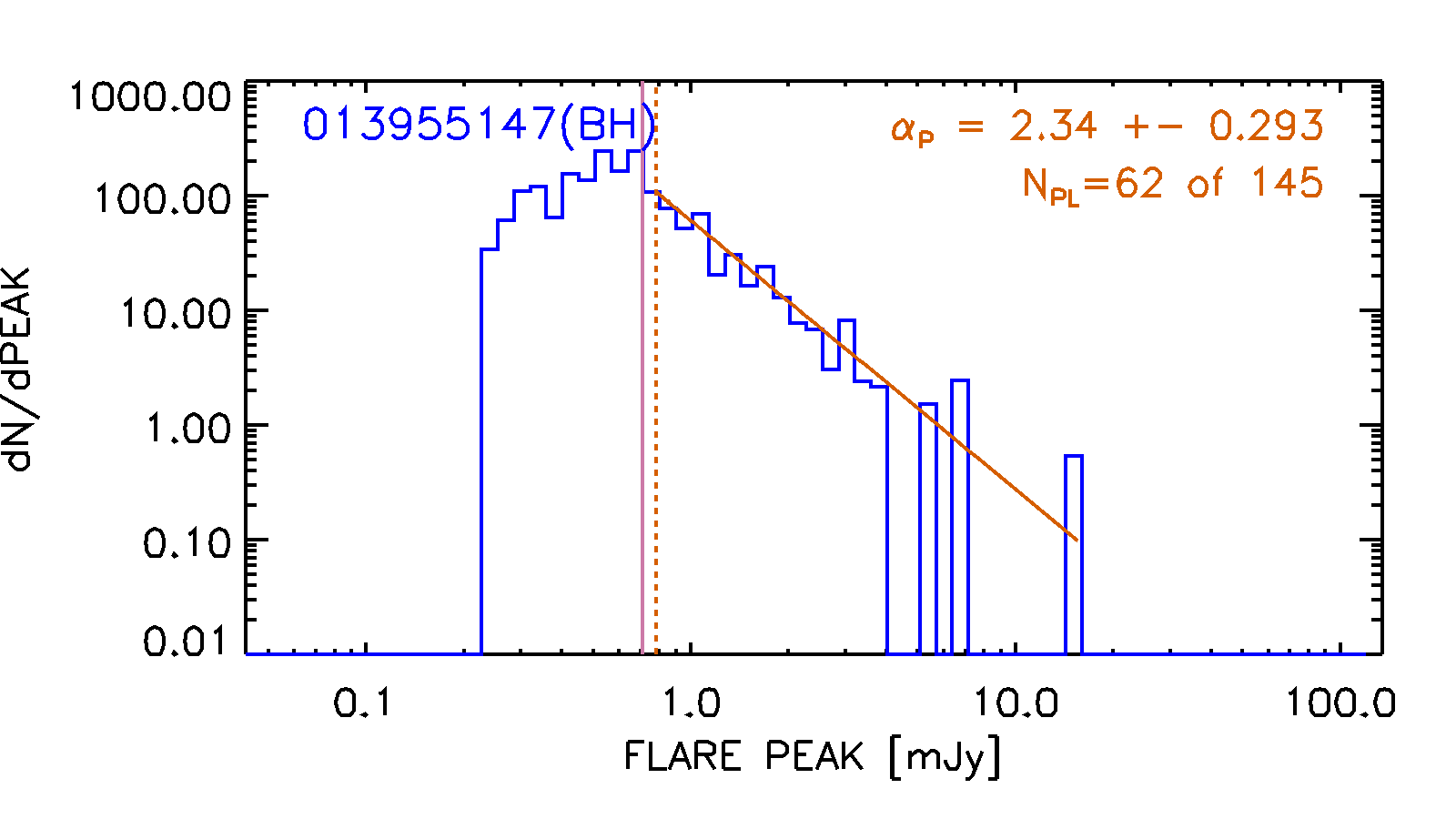}
    \includegraphics[width=0.48\linewidth]{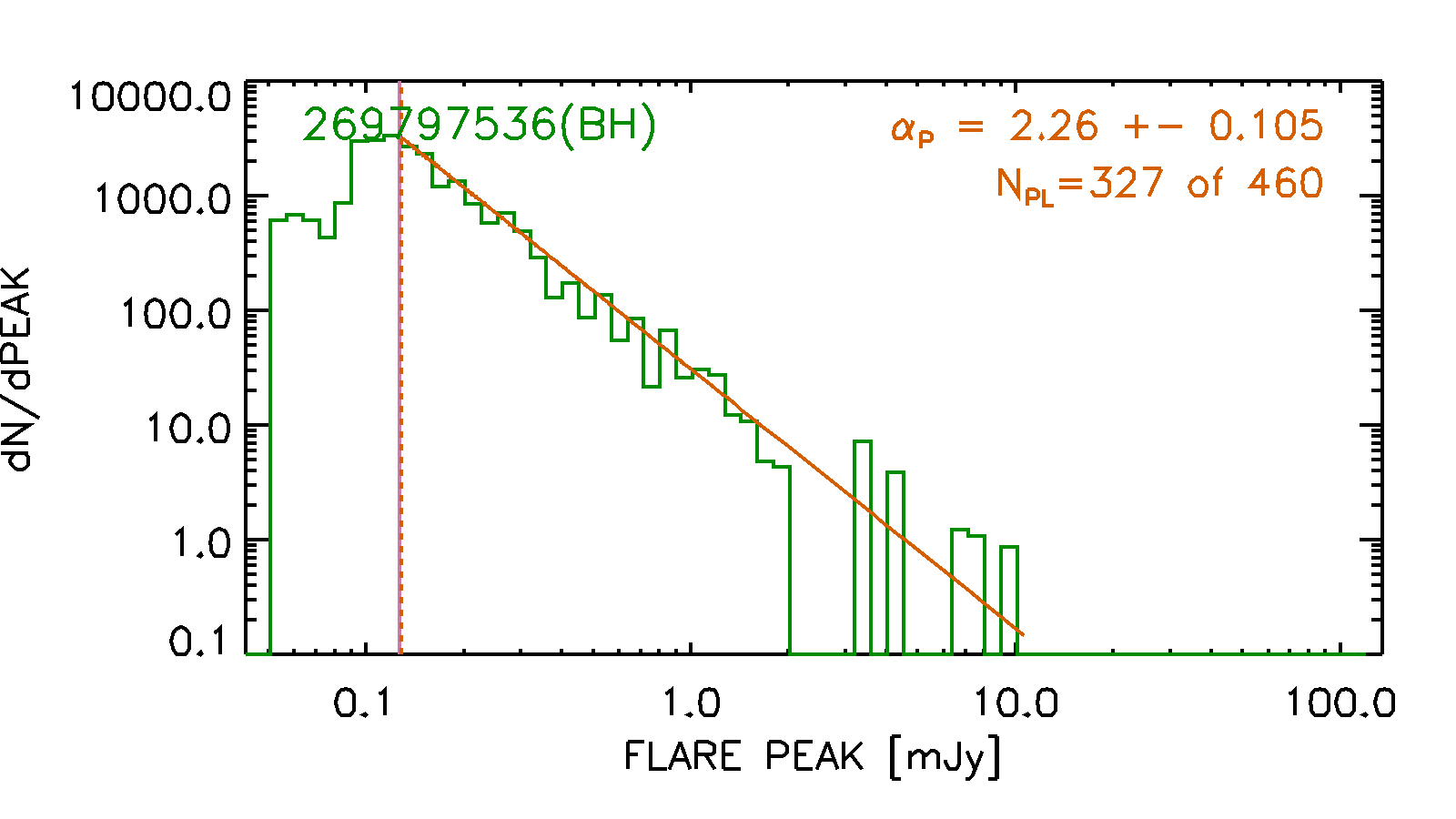}
    \caption{As in Figure~\ref{fig:dNdE}, for the peak flux of detected flares, $\frac{dN}{d\mathrm{(Peak flux)}}$.  Unlike the distributions for energies, these have similar indices $\alpha_P$, though the distribution for \ticone\ is systematically steeper than for \tictwo.}
    \label{fig:dNdP}
\end{figure*}

Estimating the power-law indices is a difficult analysis problem because flare detection efficiency decreases as flare strengths decrease, which leads to the distributions turning over at lower values of $Z$.  This will tend to make the fitted power-laws systematically flatter (i.e., result in smaller values of $\alpha_{Z}$) if the domain of the fit extends too low.  In order to ameliorate this problem, we first sort the $N$ values for each of the detected flares $\{Z_k, k=1{\ldots}N\}$ in ascending order, and compute $\{\alpha_{Z_j}, \forall Z{\ge}Z_j\}$ using the maximum likelihood method of \cite{1970ApJ...162..405C}.  For each fit, we construct cumulative distribution functions and determine the distance $d_{Z_j}$ between the empirical distribution and the model.  We expect there to be a minimum as the model fit improves as the lower bound of the fit moves into the regime where the power-law holds, until the sample sizes start decreasing and $d_{Z_j}$ begins to increase.  This is indeed the behavior we find in practice, and we choose the value of $\alpha_{Z_j}$ where $d_{Z_j}$ is the minimum as the representative value.  The empirical distributions and the model fits for the full set of detected flares (under both the BH and HB thresholding schemes) are shown in Figures~\ref{fig:dNdE} and \ref{fig:dNdP}.  The solid vertical lines indicate the mode of the $\log{Z}$ sample (computed from the sample of $Z$ as in \cite{2006ApJ...652..610P}, and the lower bound of the fit which minimizes $d_{Z}$ are marked as dashed vertical lines.

\begin{table*}
    \centering
    \caption{{Characterizing the power-law distributions for flare energy, peak flux, and impulse.  The power-law indices $\alpha_Z$, the level at which the power-law is expected to be complete, and the number of flares used in each fit are shown for each source and for each method of setting the threshold.}}
    \begin{tabular}{l | c c | c c | c c}
    \hline\hline
    Property & \multicolumn{2}{c}{\ticone}\ & \multicolumn{2}{c}{\tictwo} & \multicolumn{2}{c}{\ticfour\ (\aumic)}\\
    \hfil & (HB) & (BH) & (HB) & (BH) & (HB) & (BH) \\
    \hline
    Energies PL index $\alpha_E$ & $1.64 \pm 0.28$ & $1.87 \pm 0.20$ & $2.06 \pm 0.14$ & $2.07 \pm 0.11$ & $1.57 \pm 0.13$ & $1.76 \pm 0.11$ \\
    PL completeness ${E_\textrm{min}}$ [$10^{-9}$~erg cm$^{-2}$] & $8.1$ & $6.2$ & $3.1$ & $2.2$ & $39.5$ & $31.0$ \\
    Number of flares in PL($E$) fit & 58 & 97 & 154 & 232 & 98 & 166 \\
    \hline
    Peak flux PL index $\alpha_P$ & $2.41 \pm 0.46$ & $2.34 \pm 0.29$ & $2.15 \pm 0.15$ & $2.26 \pm 0.11$ & $1.83 \pm 0.20$ & $1.85 \pm 0.22$ \\
    PL completeness ${P_\textrm{min}}$ [mJy] & $1.27$ & $0.78$ & $0.21$ & $0.13$ & $9.74$ & $4.45$ \\
    Number of flares in PL($P$) fit & 32 & 62 & 158 & 327 & 95 & 84 \\
    \hline
    {Impulsive flux PL index $\alpha_I$} & $2.36 \pm 0.46$ & $2.48 \pm 0.34$ & $2.18 \pm 0.16$ & $2.21 \pm 0.16$ & $2.15 \pm 0.25$ & $2.24 \pm 0.20$ \\
    {PL completeness ${I_\textrm{min}}$ [mJy]} & $1.06$ & $0.79$ & $0.21$ & $0.21$ & $9.31$ & $7.15$ \\
    {Number of flares in PL($I$) fit} & 35 & 58 & 157 & 160 & 84 & 125 \\
    \hline
    \end{tabular}
    \label{tab:alphas}
\end{table*}

The results of fitting power-laws to integrated flare energies ($Z=E$), flare peak fluxes ($Z=P$), {and impulsive flux ($Z=I$)} are listed in Table~\ref{tab:alphas}.  The error bars on the $\alpha_Z$ are computed via bootstrapping.  Of note is that the $\alpha_Z$ for both BH- and HB-derived flare samples are consistent with each other in all cases, despite the samples having different lower bounds and sample sizes.  
{However, $\alpha_E$ and $\alpha_P$ show significant differences between the different stars. For the solar-like star \ticone, $\alpha_E<2$.  For the low-mass M dwarfs, $\alpha_E<2$ for the young active flare star \ticfour\ (\aumic), but $\alpha_E>2$ for the less active M dwarf \tictwo.  The peak flux distribution is flatter for the highly active \ticfour\ ($\alpha_P<2$, while it is steeper for the less active stars \ticone\ and \tictwo\ ($\alpha_P>2$).  The solar-like star appears to have steepest distribution, though the uncertainties are large enough to preclude statistical significance.  In contrast, $\alpha_I>2$ systematically for all the stars, and is statistically indistinguishable.  Taken at face value, this suggests that the physics of flare {\sl onset} is the same in all the stars, but the subsequent response of the corona is environment dependent.}

{The estimated values of $\alpha_E$ and $\alpha_P$ for the solar-like \ticone\ are similar to those observed on the Sun \citep[][]{2012ApJ...754..112A} in X-rays.}  The large increase in the steepness of the distributions between $\alpha_E$ and $\alpha_P$ for \ticone\ does suggest that either less energetic flares produce higher peak fluxes or that more energetic flares produce lower peak fluxes compared to \tictwo.  We also report the minimum values we derive for the applicability of the power-law fits, and interpret them as a measure of the statistical completeness of the sample: that is, we expect that the intrinsic flare distribution continues to be described by the same power-law at lower $Z$, but that the observed numbers fall off due to detection sensitivity.

\subsection{{Advantages, Comparisons, Limitations}}\label{sec:compare}

{Our method has three unique characteristics: first, we use a flexible, yet physically motivated, harmonic detrending to robustly locate the baseline; second, we locate flares as large deviations that cannot be accounted for with the general correlation structure in the detrended residuals; and we control for false discovery rates and the family-wise error rates.  These allow us to detect even weaker flares reliably even under challenging conditions like sharply changing baselines or large flaring rates, allowing us to build a more complete and representative population of flares.}

{We expect our detected samples to be complete at the 50\% level to $\approx0.1$\% of peak amplitude relative to the estimated bolometric luminosity for \ticone, to $\approx0.6$\% for \tictwo, and to $\approx0.2$\% for \ticfour\ (\aumic); and the smallest flare amplitudes we detect are at $0.03-0.05$\%, $\approx0.3$\%, and $0.06-0.007$\% of $L_\textrm{bol}$ for each star respectively; the last of which is comparable to the smallest flare amplitude detected in \aumic\ by \citet{2021A&A...649A.177M}.}

{The smallest flare energies detected at the permissive BH threshold are $7.9\times10^{32}$~erg for \ticone, $7.6\times10^{31}$~erg for \tictwo, and $3.2\times10^{29}$~erg for \ticfour\ (\aumic).  This is comparable to that found in 2~s cadence data of \aumic\ by \citet{2022AJ....163..147G}, and an order of magnitude lower than that found in 2~min cadence data by both \citealt{2022A&A...661A.148C} and \citet{2022AJ....163..147G}.  Note that in their comprehensive analysis of TESS M dwarfs, \citet{2025arXiv250515451G} find, for some representative stars (BD-15\,6290, YZ\,CMi, EV\,Lac, all of which are at $\approx5$~pc) the smallest flare energies at $\approx2-3\times10^{-10}$~erg~cm$^{-2}$; we reach a significantly deeper sensitivity of $3\times10^{-11}$~erg~cm$^{-2}$ for \aumic, even with its farther distance, with our method.
}

{As can be expected from the improved sensitivity, we also improve upon the total number of detected flares.  In \aumic, we detect 149 flares under the HB threshold, and as many as 402 under the BH threshold, compared to $\approx100$ flares found by \citet{2022AJ....163..147G}, $\approx250$ flares found by \citet{2022A&A...661A.148C}, and $324$ flares found by \citet{2021A&A...649A.177M}.}

Our goal is to eventually carry out a systematic detection and characterization of $\alpha_Z$ across the main sequence, in a manner pioneered by \cite{Feinstein2022} and \cite{2025arXiv250515451G}.  As noted in these studies, the measurement of the slopes of the flare frequency distributions is a diagnostic of self-organized criticality (SOC), which is thought to be the driver for flare onset in solar and stellar coronae \citep[][]{1991ApJ...380L..89L}.  Our method will then provide robust improvements in assessments of the baseline, flare detection, as well as {in lowering the thresholds at which the the flare frequency distributions show turnovers \citep[e.g.,][]{2022A&A...661A.148C,2025arXiv250515451G}.}
%excluding the inevitable turnovers in the flare frequency distributions.  
{We note that in the exemplar cases we study here, the solar-like star (\ticone) behaves as in the Sun, with $\alpha_E\approx1.8$ and $\alpha_P\gtrsim2$.  For the less active low mass star (\tictwo), $\alpha_E$ and $\alpha_P$ are significantly steeper than for equivalent stellar types in \cite{Feinstein2022}, which we attribute to a better assessment of the lower bound of applicability of a power-law distribution.}  In concurrence with several studies \citep[e.g.,][]{2000ApJ...541..396A,2025arXiv250515451G}, we find that late type stars can indeed have distributions with steep $\alpha_E$.
%Since our target stars are ${\approx}15\times$ more distant than the cases highlighted by \cite{2025arXiv250515451G}, the break in the power-law they report is below the detection limit for \tictwo.
{In contrast, the young and highly active flare star \aumic\ has significantly flatter slopes, with both $\alpha_E,\alpha_P<2$.}

{An important requirement for our analysis is that there exist sufficient non-flare signal in the light curve for the GARCH analysis to work.  If the light curve has too many flares, the signatures of the flares will be subsumed into the residuals, and individual flares would not be detected.  This may happen for $\alpha_{E,P}>>3$ when large numbers of small flares may constitute most of the observed light curve.  In all the cases considered, $85-90$\% of the fraction of the light curves are free of candidate detections.}

{Here, we ignore the effect of exoplanet transits in the light curves.  This is usually a small effect, with transit depths $\ll1$\% of the baseline.  Even when they are not modeled out by our harmonic model (see tell tale signature of the transit of \aumic\,b at time 1331~d in Figure~\ref{fig:AUMic23}~left), they only affect the distribution of $\varepsilon_t$ over the negative deviations, and do not affect flare detections.}

{As discussed in Section~\ref{PFD}, our detection method works by modeling the time-lag correlations in the light curve, and is thus sensitive to the instances when impulses occur that lead to flares.  This method may thus provide a window into the study of flare onset.  We model the distribution of the impulses (defined by $Z_t$ at the flare detection times) as power-laws and report the results in Table~\ref{tab:alphas}.  The estimated power-law indices are all remarkably similar, with $\alpha_I$ consistent with the range $2.2-2.3$, suggesting that the physics of flare onset is the same across the program stars, even if their response to the flares may depend on the particular environments of the respective coronae.}

\section{Summary}\label{sec:summary}

We have developed a new method to analyze TESS time series to detect flares in the presence of an oscillatory baseline emission.  We introduce a novel two-stage statistical model that combines a deterministic trend with stochastic volatility (see Section~\ref{sec:model}).  We first model and remove the underlying non-stochastic trend using a harmonic fit with time-varying amplitudes and phases that is capable of modeling complex variability arising from stellar surface variability (see Equation~\ref{eqn:harmonic_amplitudes}).  We then model the residuals using a stochastic volatility model analogously to financial time series, by fitting to an ARMA and GARCH model (see Equations~\ref{eq:Z:specification}-\ref{eq:arma:model}).  We construct thresholds for flare detection in a rigorous way, using two different techniques based on the False Discovery Rate and the Family-wise Error Rate (see Section~\ref{sec:HB_BH}), and demonstrate using simulations (see Section~\ref{sec:simulation}) that stellar flares can indeed be detected as significant deviates from the baseline.

We apply this method to several TESS light curves from different stars (see Table~\ref{tab:stars}) observed at different cadences, and show that hundreds of flares are detected (see Table~\ref{tab:flarecensus}) down to a minimum peak flux of $\approx10^{-13}$~erg~s$^{-1}$~cm$^{-2}$ (Footnote~\ref{foot:rate2fx}). %0.06~mJy.  
Typical flare rates range from $\approx$0.4-2.7~day$^{-1}$ depending on the star, the particular observation, and the adopted detectability thresholds.  We fit power-law distributions to both the flare energies and peak fluxes (see Table~\ref{tab:alphas}), and find that the former are different for the two lower-activity stars, with the solar-like star exhibiting an index $\alpha_E<2$ (similar to that seen in the Sun) and the late type star exhibiting $\alpha_E\gtrsim2$ (consistent with other estimates for late-type stars), while the latter show $\alpha_P>2$ for both stars.  This difference is suggestive of fundamental differences in the heating mechanisms of the two stars, and a comprehensive analysis of a large sample of TESS observations is needed to establish whether there are discernible trends with stellar properties.

\section*{Acknowledgements}

This paper includes data collected by the TESS mission, which are publicly available from the Mikulski Archive for Space Telescopes (MAST). Funding for the TESS mission is provided by NASA's Science Mission directorate.
This research has made use of the SIMBAD database,
operated at CDS, Strasbourg, France.
We thank members of the CHASC collaboration for helpful discussions.  We thank Gwen Eadie and David van Dyk for useful comments on detrending and the multiple testing problem, Hyungsuk Tak as the inspiration behind Figure~\ref{fig:bh_hb_thr}, and Eric Feigelson for useful discussions on the nature and applicability of ARMA processes in astronomy.  We thank the anonymous referee for comments and suggestions that helped to improve the paper.  V.L.K.\ was supported during this work by the NASA Contract to the Chandra X-ray Center AR3-24002X.  G.M.\ and Q.J.W.\ acknowledge support by Columbia University, and Q.J.W.\ further acknowledges support by Texas A\&M University.

% The Acknowledgements section is not numbered. Here you can thank helpful
% colleagues, acknowledge funding agencies, telescopes and facilities used etc.

%%%%%%%%%%%%%%%%%%%%%%%%%%%%%%%%%%%%%%%%%%%%%%%%%%
\section*{Data Availability}

TESS data used in this work is available at \href{https://archive.stsci.edu/doi/resolve/resolve.html?doi=10.17909/1kjm-5506}{DOI:10.17909/1kjm-5506}.
The figures generated for the rest of the datasets are available at the Zenodo repository \url{https://doi.org/10.5281/zenodo.14991246} \citep{wang_2025_14991246}.
The Python code implementing the flare detection algorithm and the example TESS light curve data analyzed in this study are publicly available at \url{https://github.com/johnnywalker213/Detecting-stellar-flares}.
% The inclusion of a Data Availability Statement is a requirement for articles published in MNRAS. Data Availability Statements provide a standardised format for readers to understand the availability of data underlying the research results described in the article. The statement may refer to original data generated in the course of the study or to third-party data analysed in the article. The statement should describe and provide means of access, where possible, by linking to the data or providing the required accession numbers for the relevant databases or DOIs.



\bibliographystyle{mnras}
\bibliography{references} % if your bibtex file is called references.bib

%\begin{thebibliography}{99}
%\end{thebibliography}

%%%%%%%%%%%%%%%%%%%%%%%%%%%%%%%%%%%%%%%%%%%%%%%%%%

%%%%%%%%%%%%%%%%% APPENDICES %%%%%%%%%%%%%%%%%%%%%

\appendix
\section{AU Mic Analysis}\label{app:aumic}

We show figures for the results of the analysis of \ticfour\ (\aumic) here, corresponding to equivalent analyses of \ticone\ and \tictwo\ with Figures~\ref{fig:harmonic_fitting}, \ref{fig:quatile_detect_demo}, \ref{fig:epsilont_dist}, \ref{fig:bh_hb_thr}, \ref{fig:dNdE}, and \ref{fig:dNdP}.

\begin{figure*}
    \centering
    \includegraphics[width=0.45\linewidth]{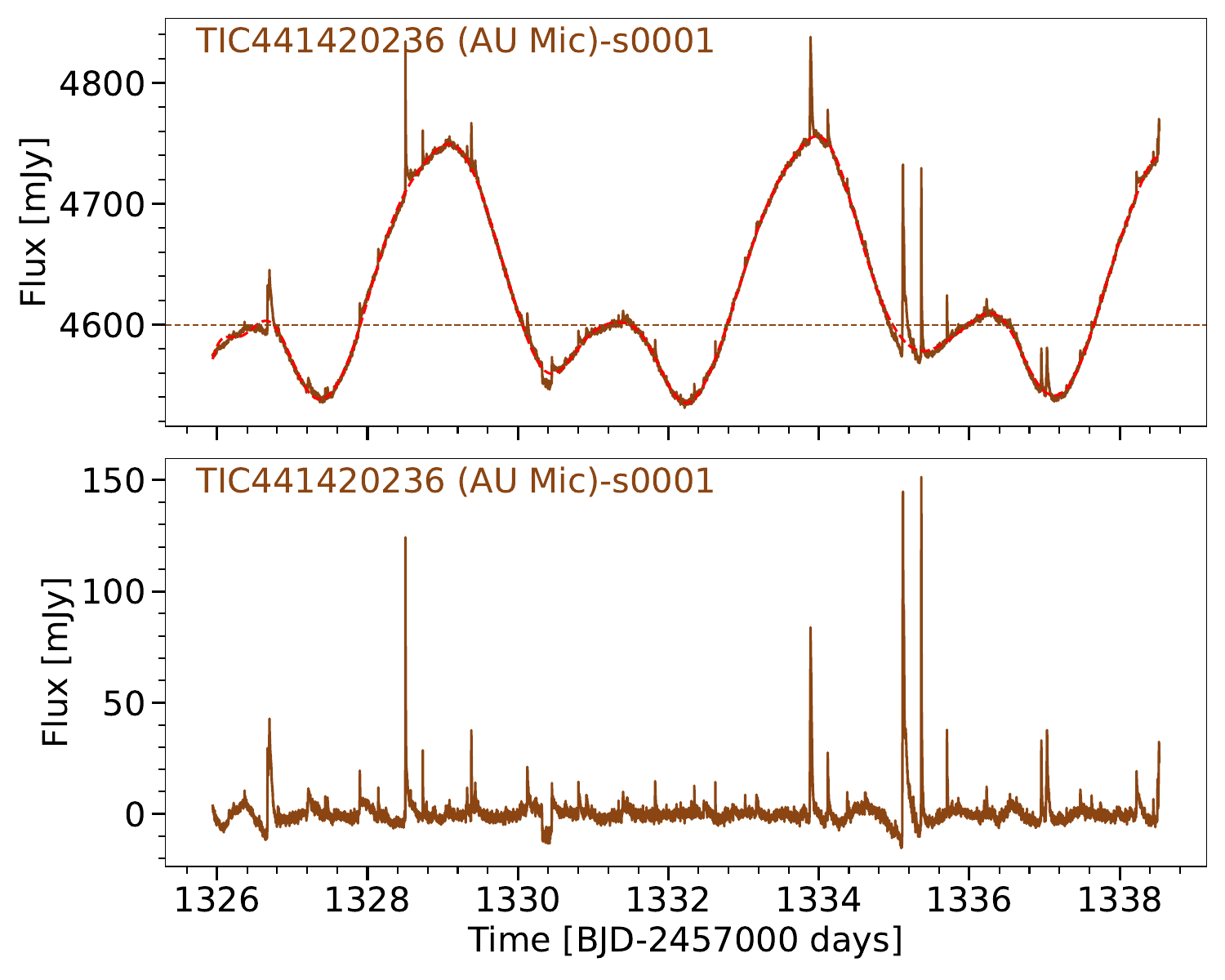}
    \includegraphics[width=0.45\linewidth]{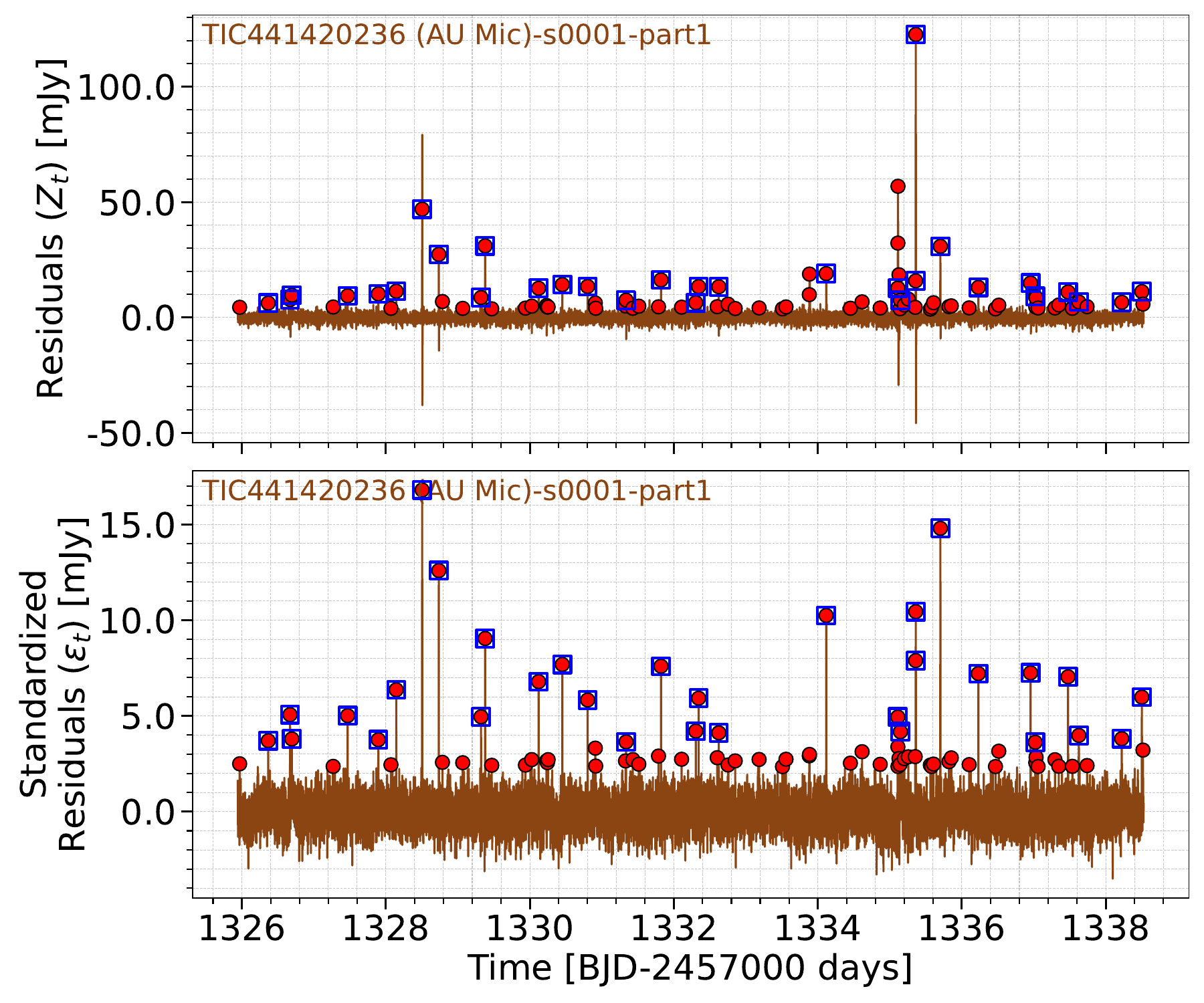}
    \caption{Analysis of the first part of sector 1 of \ticfour\ (\aumic), as in Figure~\ref{fig:harmonic_fitting} (left), and as in Figure~\ref{fig:quatile_detect_demo} (right).}
    \label{fig:AUMic23}
\end{figure*}

\begin{figure*}
    \centering
    \includegraphics[width=0.45\linewidth]{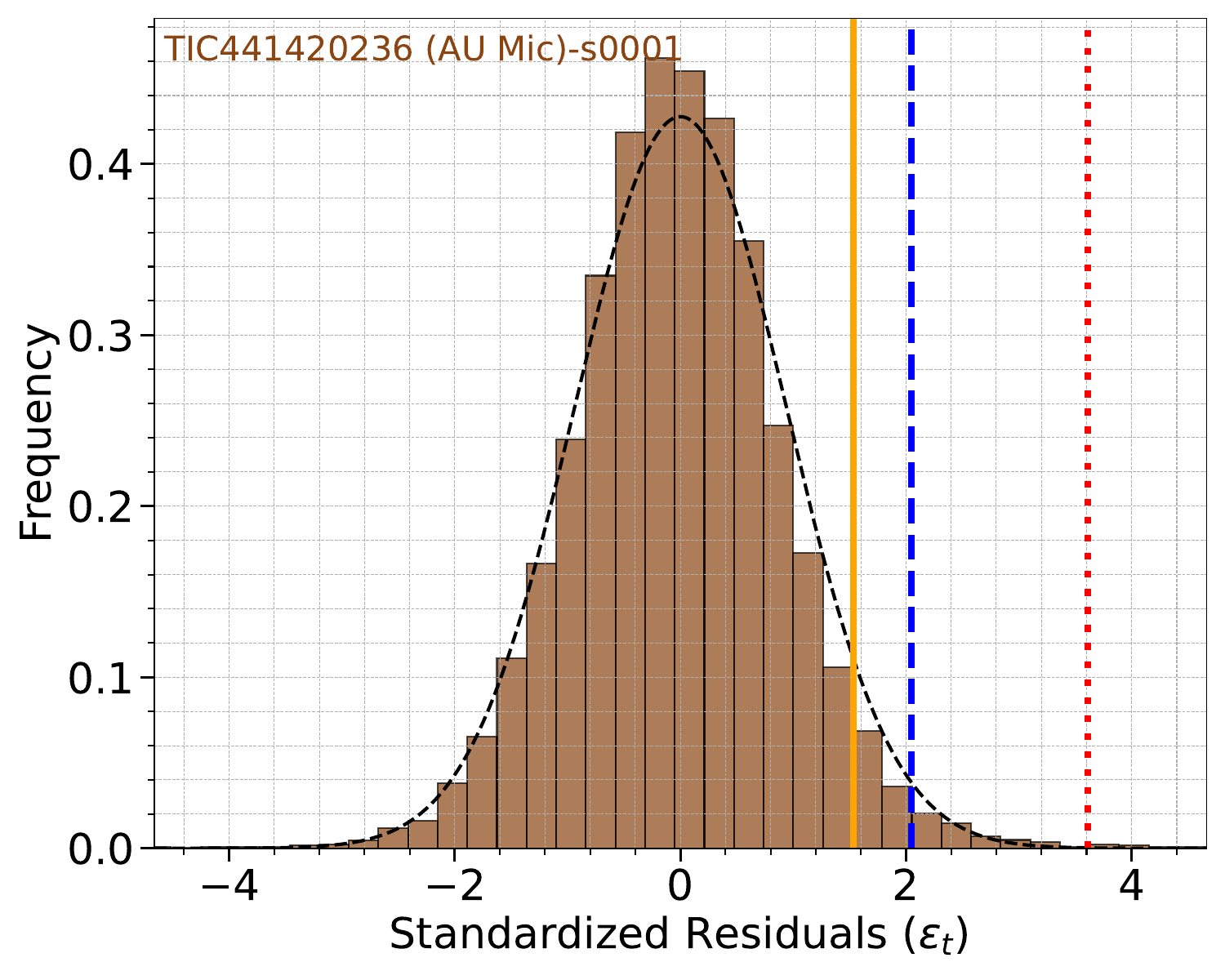}
    \includegraphics[width=0.45\linewidth]{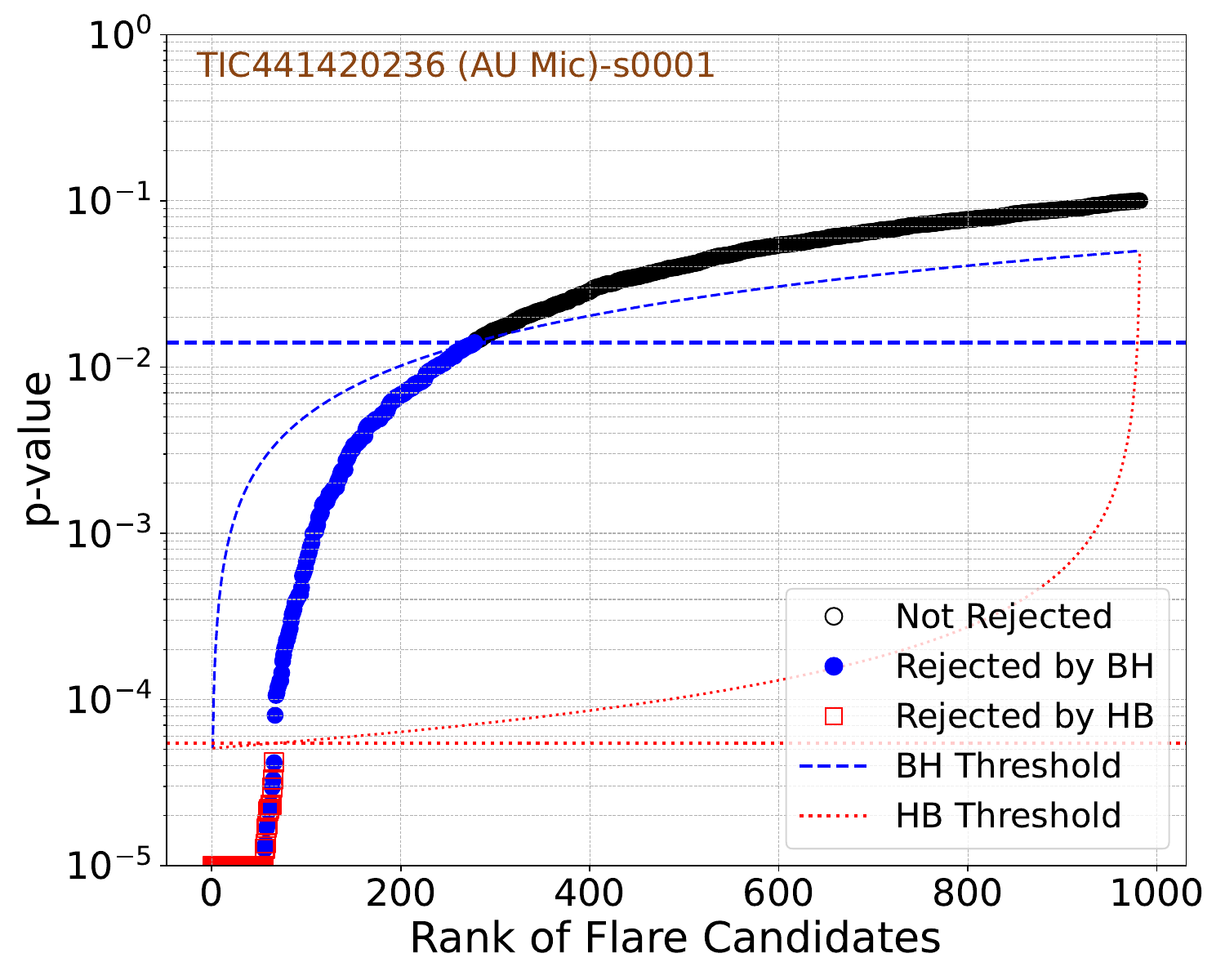}
    \caption{Flare detections in sector 1 of \ticfour\ (\aumic), as in Figure~\ref{fig:epsilont_dist} (left), and as in Figure~\ref{fig:bh_hb_thr} (right).}
    \label{fig:AUMic45}
\end{figure*}

% \begin{figure*}
%     \centering
%     \includegraphics[height=2.8in]{TIC_441420236_s0001_1_deharmonic.pdf}
%     \includegraphics[height=2.8in]{TIC_441420236_s0027_1_deharmonic.pdf}
%     \caption{
%     As in Figure~\ref{fig:harmonic_fitting}, for \ticfour\ (\aumic).
%     }
%     \label{fig:AU_Mic_harmonic_fitting}
% \end{figure*}

% \begin{figure*}
% \centering
%     \includegraphics[height=2.8in]{TIC_441420236_s0001_1_flares_analysis.pdf}
%     \includegraphics[height=2.8in]{TIC_441420236_s0027_1_flares_analysis.pdf}
%     \caption{
%     As in Figure~\ref{fig:quatile_detect_demo}, for \ticfour\ (\aumic).
%     }
% \label{fig:AU_Mic_quatile_detect_demo}
% \end{figure*}

% \begin{figure*}
% \centering
%     \includegraphics[height=2.8in]{TIC441420236_s0001_epsilon_distribution.pdf}
%     \includegraphics[height=2.8in]{TIC441420236_s0027_epsilon_distribution.pdf}
%     \caption{
%     As in Figure~\ref{fig:epsilont_dist}, for \ticfour\ (\aumic).
%     }
% \label{fig:AU_Mic_epsilont_dist}
% \end{figure*}

% \begin{figure*}
% \centering
%     \includegraphics[height=2.7in]{TIC441420236_s0001_updated_threshold_plot.pdf}
%     \includegraphics[height=2.7in]{TIC441420236_s0027_updated_threshold_plot.pdf}
% \caption{As in Figure~\ref{fig:bh_hb_thr}, for \ticfour\ (\aumic).
% }
% \label{fig:AU_Mic_bh_hb_thr}
% \end{figure*}

\begin{figure}
    \centering
    \includegraphics[width=\linewidth]{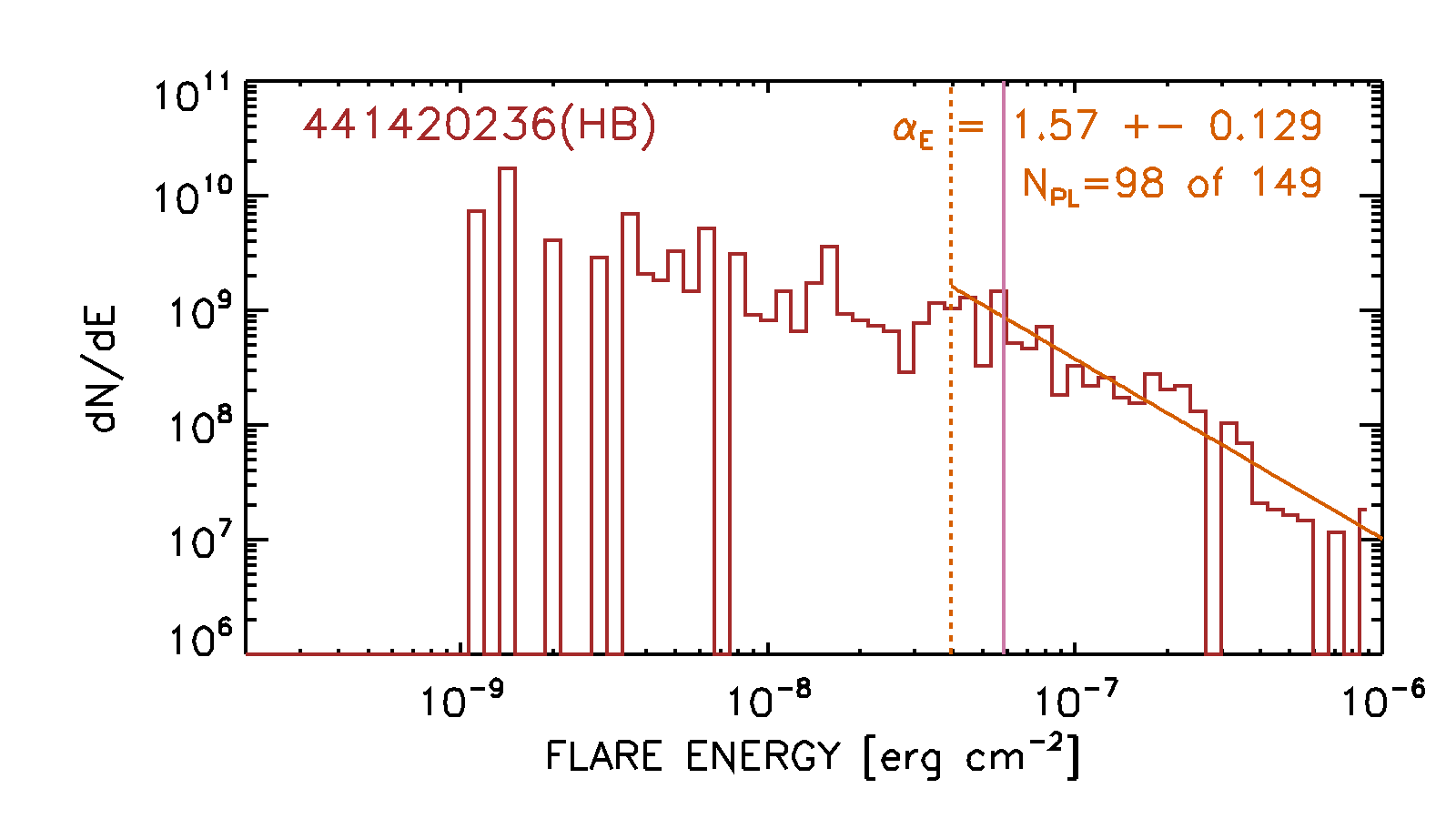}
    \includegraphics[width=\linewidth]{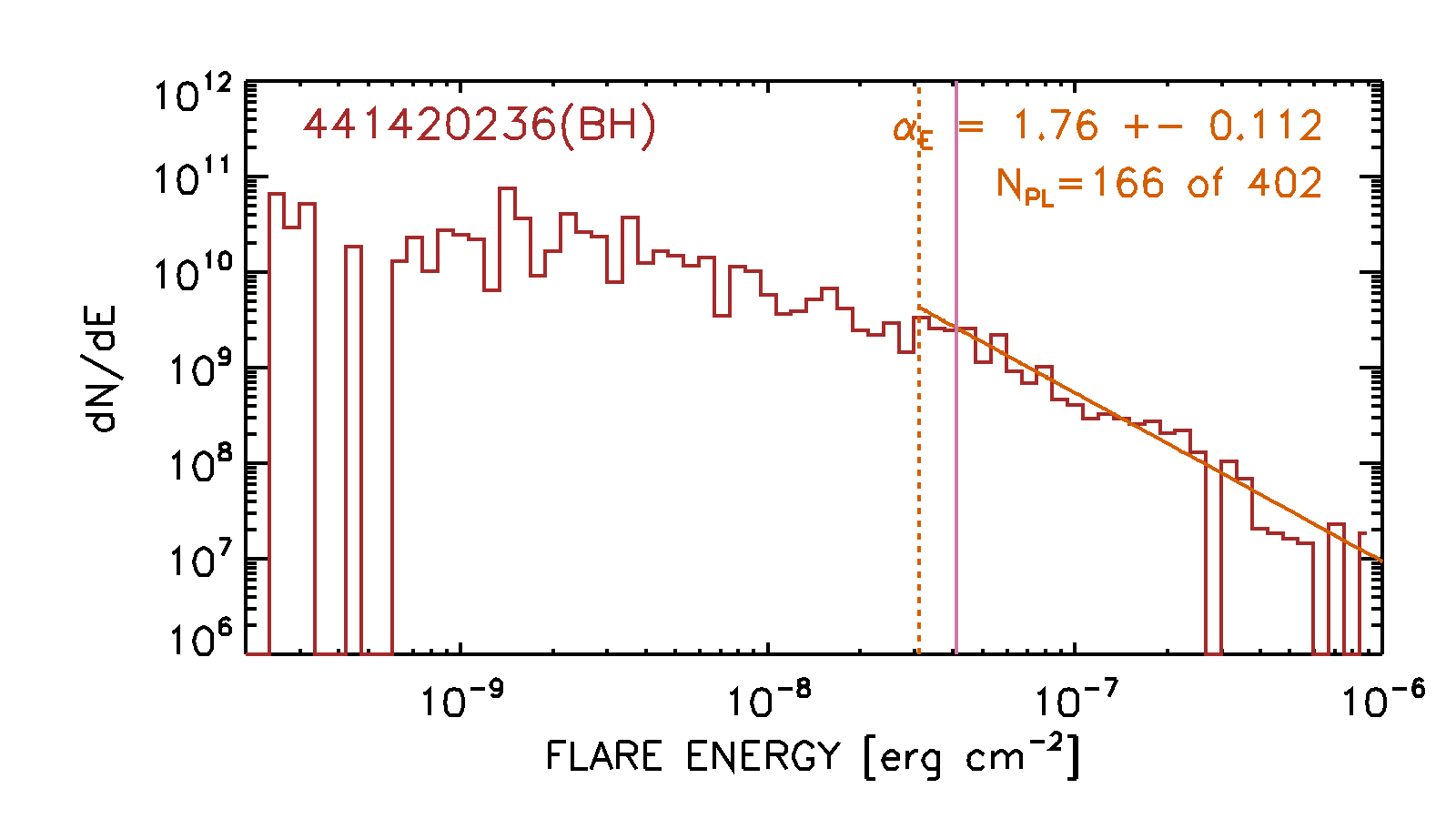}
    \caption{As in Figure~\ref{fig:dNdE}, for \ticfour\ (\aumic).}
    \label{fig:dNdE4}
\end{figure}

\begin{figure}
    \centering
    \includegraphics[width=\linewidth]{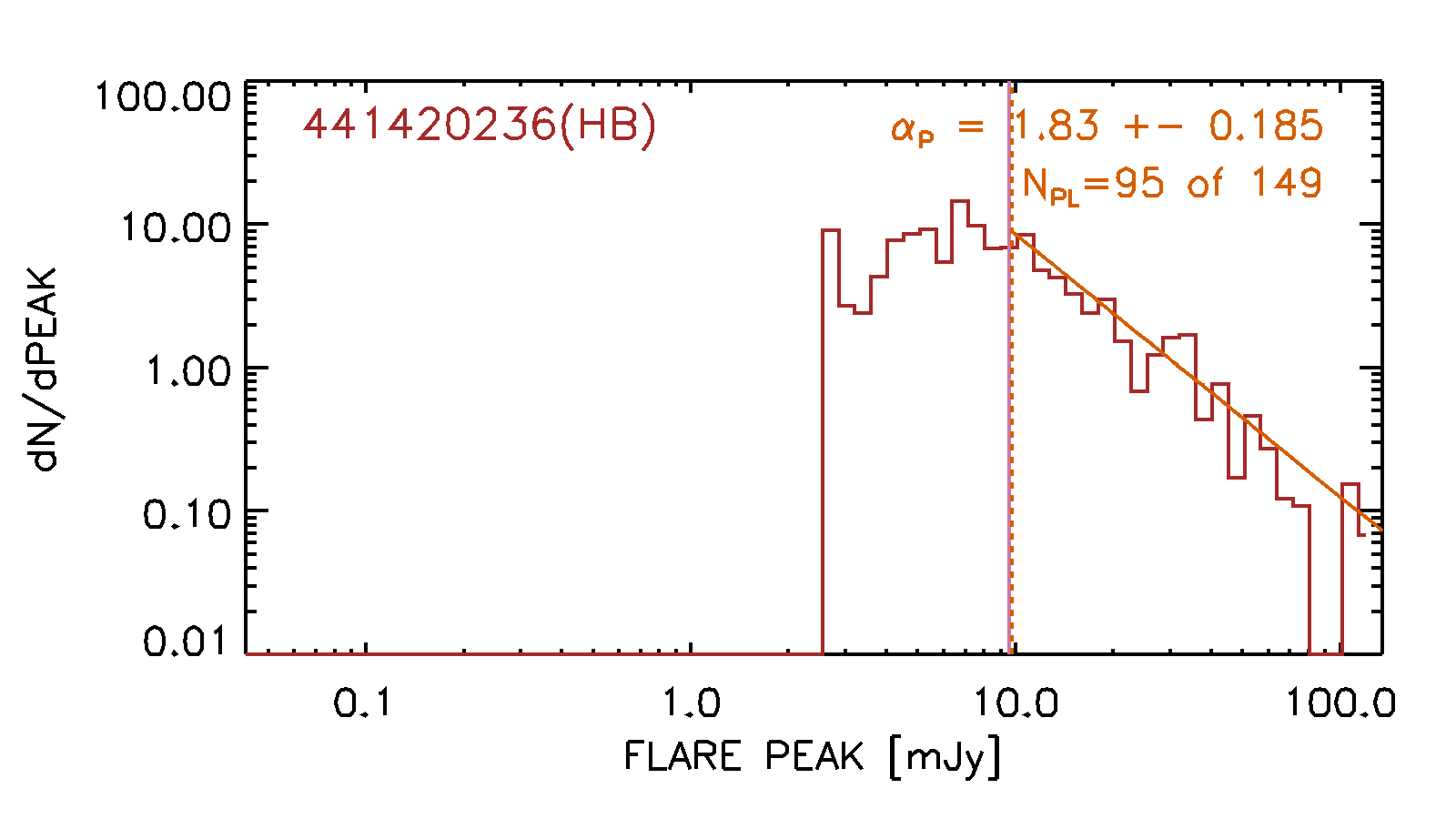}
    \includegraphics[width=\linewidth]{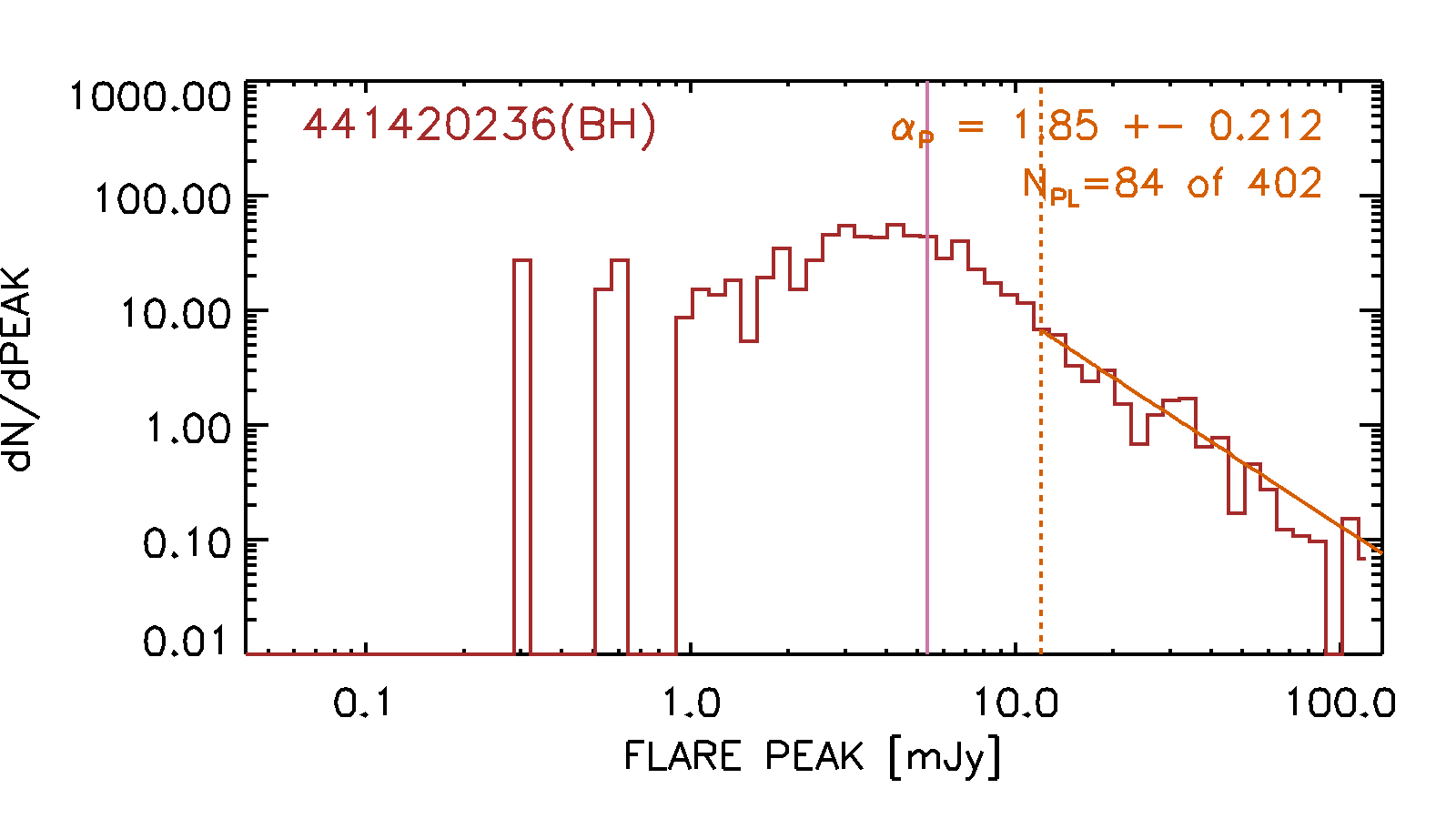}
    \caption{As in Figure~\ref{fig:dNdP}, for \ticfour\ (\aumic).}
    \label{fig:dNdP4}
\end{figure}

% Don't change these lines
\bsp	% typesetting comment
\label{lastpage}
\end{document}